\documentclass[aps,prd,superscriptaddress,showpacs,preprintnumbers,floatfix]{revtex4}
\usepackage{graphicx,color}
\newcommand{\bm}[1]{\mbox{\boldmath $#1$}}
\newcommand{\be}{\begin{equation}}
\newcommand{\ee}{\end{equation}}
\newcommand{\bea}{\begin{eqnarray}}
\newcommand{\eea}{\end{eqnarray}}

\newcommand{\bfk}{\mbox{\boldmath $k$}}

\newcommand{\bfp}{\mbox{\boldmath $p$}}

\newcommand{\bfP}{\mbox{\boldmath $P$}}

\newcommand{\qup}{q^\uparrow}

\newcommand{\ua}{\uparrow}

\def\lsim{\mathrel{\rlap{\lower4pt\hbox{\hskip1pt$\sim$}}\raise1pt\hbox{$<$}}}
\def\gsim{\mathrel{\rlap{\lower4pt\hbox{\hskip1pt$\sim$}}\raise1pt\hbox{$>$}}}
\def\nostrocostruttino#1\over#2{\mathrel{\mathop{\kern 0pt \rlap
{\hbox{$#1$}}} \hbox{\kern-.135em $#2$}}}

\def\kt{k_\perp}

\def\bkt{\bfk_\perp}

\def\pp{p_\perp}
\def\bpp{\bfp_\perp}

\def\xb{x_{_{\!B}}}

\def\avk{\langle k_\perp ^2\rangle}
\def\avp{\langle p_\perp ^2\rangle}
\def\avPT{\langle P_T^2\rangle}

\def\T{_{_T}}
\def\C{_{_C}}

\begin{document}
\preprint{JLAB-THY-13-1704} 
\title{Simultaneous extraction of transversity and Collins functions \\ 
from new SIDIS and $e^+e^-$ data}
\author{M.~Anselmino}
\affiliation{Dipartimento di Fisica, Universit\`a di Torino,
             Via P.~Giuria 1, I-10125 Torino, Italy}
\affiliation{INFN, Sezione di Torino, Via P.~Giuria 1, I-10125 Torino, Italy}
\author{M.~Boglione}
\affiliation{Dipartimento di Fisica, Universit\`a di Torino,
             Via P.~Giuria 1, I-10125 Torino, Italy}
\affiliation{INFN, Sezione di Torino, Via P.~Giuria 1, I-10125 Torino, Italy}
\author{U.~D'Alesio}
\affiliation{Dipartimento di Fisica, Universit\`a di Cagliari,
             Cittadella Universitaria, I-09042 Monserrato (CA), Italy}
\affiliation{INFN, Sezione di Cagliari,
             C.P.~170, I-09042 Monserrato (CA), Italy}
\author{S.~Melis}
\affiliation{Dipartimento di Fisica, Universit\`a di Torino,
             Via P.~Giuria 1, I-10125 Torino, Italy}
\affiliation{INFN, Sezione di Torino, Via P.~Giuria 1, I-10125 Torino, Italy}
\author{F.~Murgia}
\affiliation{INFN, Sezione di Cagliari,
             C.P.~170, I-09042 Monserrato (CA), Italy}
\author{A.~Prokudin}
\affiliation{Jefferson Laboratory, 12000 Jefferson Avenue, Newport News, VA 23606, USA}
\date{\today}

\begin{abstract}

We present a global re-analysis of the most recent experimental
data on azimuthal asymmetries in semi-inclusive deep inelastic
scattering, from the HERMES and COMPASS Collaborations, and
in $e^+e^- \to h_1 \, h_2 \, X$ processes, from the Belle Collaboration. 
The transversity and the Collins functions are extracted simultaneously, 
in the framework of a revised analysis in which a new 
parameterisation of the Collins functions is also tested. 
\end{abstract}


\maketitle

\section{\label{Intro} Introduction and formalism}

The spin structure of the nucleon, in its partonic collinear configuration,
is fully described, at leading-twist, by three independent Parton Distribution Functions (PDFs): the unpolarised PDF, the helicity distribution and 
the transversity distribution. While the unpolarised PDF and the helicity distribution, which have been studied for decades, are by now very 
well or reasonably well known, much less information is available on 
the latter, which has been studied only recently. The reason is that, due 
to its chiral-odd nature, a transversity distribution can only be accessed 
in processes where it couples to another chiral-odd quantity. 

The chiral-odd partner of the transversity distribution could be a 
fragmentation function, like the Collins function~\cite{Collins:1992kk}
or the di-hadron fragmentation function~\cite{Collins:1993kq, Jaffe:1997hf,
Radici:2001na} or another parton distribution, like the 
Boer-Mulders~\cite{Boer:1997nt} or the transversity distribution itself.     
A chiral-odd partonic distribution couples to a chiral-odd fragmentation
function in Semi-Inclusive Deep Inelastic Scattering processes (SIDIS, 
$\ell \, N \to \ell \, h \, X$). The coupling of two chiral-odd partonic 
distributions could occur in Drell-Yan processes (D-Y, $p \, N \to \ell^+ 
\ell^- X$) but, so far, no data on polarised D-Y is available. 
Information on the convolution of two chiral-odd fragmentation functions 
(FFs) can be obtained from $e^+e^- \to h_1 \, h_2 \, X$ processes.    

The $u$ and $d$ quark transversity distributions, together with the Collins 
fragmentation functions, have been extracted for the first time in 
Refs.~\cite{Anselmino:2007fs, Anselmino:2008jk}, from a combined analysis 
of SIDIS and $e^+e^-$ data. Similar results on the transversity distributions, 
coupled to the di-hadron, rather than the Collins, fragmentation function,
have been obtained recently~\cite{Bacchetta:2012ty}. These independent results 
establish with certainty the role played by the transversity distributions in 
SIDIS azimuthal asymmetries.       

Since the first papers~\cite{Anselmino:2007fs, Anselmino:2008jk}, 
new data have become available: from the COMPASS experiment operating on 
a transversely polarised proton 
(NH$_3$ target)~\cite{Adolph:2012sn,Martin:2013eja}, 
from a final analysis of the HERMES Collaboration~\cite{Airapetian:2010ds} 
and from corrected results of the Belle Collaboration~\cite{Seidl:2012er}. 
This fresh information motivates a new global analysis for the simultaneous extraction of the transversity distributions and the Collins functions. 

This is performed using techniques similar to those implemented in 
Refs.~\cite{Anselmino:2007fs, Anselmino:2008jk}; in addition, a second, 
different parameterisation of the Collins function will be tested, in order 
to assess the influence of a particular functional form on our results.  
 
Let us briefly recall the strategy followed and the formalism adopted in 
extracting the transversity and Collins distribution functions from 
independent SIDIS and $e^+e^-$ data.   
 
\subsection{SIDIS}
 
We consider, at ${\cal O}(k_\perp/Q)$, the SIDIS process $\ell \, 
p^\uparrow \to \ell^\prime \, h \, X$ and the single spin asymmetry,

\be A_{UT}^{\sin(\phi_h + \phi_S)} = 2 \, \frac{\int d\phi_h \, d\phi_S \, 
[d\sigma^\uparrow - d\sigma^\downarrow] \, \sin(\phi_h +\phi_S)} 
{\int d\phi_h \, d\phi_S \, [d\sigma^\uparrow + d\sigma^\downarrow]} \>,
\label{sidis-asym}
\ee
where $d\sigma^{\uparrow,\downarrow}$ is a shorthand notation for 
$$d\sigma^{\uparrow,\downarrow} \equiv
\frac{d^6\sigma^{\ell p^{\uparrow,\downarrow} \to \ell h X }}
{dx \, dy \, dz \, d^2 \bm{P}_T \, d\phi_S}$$ 
and $x,y,z$ are the usual SIDIS variables:
\be
x = \xb = \frac{Q^2}{2(P \cdot q)} \quad\quad
y = \frac{(P \cdot q)}{(P \cdot \ell)} =\frac{Q^2}{x \, s} \quad\quad
z = z_h = \frac{(P \cdot P_h)}{(P \cdot q)} \> \cdot
\ee

We adopt here the same notations and kinematical variables as defined in 
Refs.~\cite{Anselmino:2007fs, Anselmino:2011ch}, to which we refer for 
further details, in particular for the definition of the azimuthal angles 
which appear above and in the following equations.   
 
By considering the $\sin(\phi_h+\phi_S)$ moment of 
$A_{UT}$~\cite{Bacchetta:2006tn}, we are able to single out the effect 
originating from the spin dependent part of the fragmentation function 
of a transversely polarised quark, embedded in the Collins function, 
$\Delta^N\!D_{h/q^\uparrow}(z,p_\perp) = (2 \, p_\perp/z \, m_h)\,
H_1^{\perp q}(z,\pp)$~\cite{Bacchetta:2004jz}, coupled to the TMD 
transversity distribution $\Delta_T q(x,k_\perp)$~\cite{Anselmino:2007fs}:
\be
A^{\sin (\phi_h+\phi_S)}_{UT} = \label{sin-asym}
\frac{\displaystyle  \sum_q e_q^2  \! \! \int \! \!{d\phi_h \, d\phi_S \, d^2
\bfk _\perp}\,\Delta _T q (x,\kt) \,
\frac{d (\Delta {\hat \sigma})}{dy}\,
\Delta^N D_{h/q^\ua}(z,\pp) \sin(\phi_S + \varphi +\phi_q^h)
\sin(\phi_h + \phi_S) } {\displaystyle \sum_q e_q^2 \, \int {d\phi_h
\,d\phi_S \, d^2 \bfk _\perp}\; f_{q/p}(x,k _\perp) \;
\frac{d\hat\sigma}{dy}\; \; D_{h/q}(z,p_\perp) } \>,
\ee
where $\bpp = \bfP_T - z \bkt$, and
\be
\frac{d\hat\sigma}{dy} = \frac{2\pi\alpha^2}{sxy^2} [1 + (1-y)^2]
\quad\quad\quad
\frac{d(\Delta\hat\sigma)}{dy} \equiv  
\frac{d\hat\sigma^{\ell q^\uparrow \to \ell q^\uparrow}}{dy} -
\frac{d\hat\sigma^{\ell q^\uparrow \to \ell q^\downarrow}}{dy} =
\frac{4\pi\alpha^2}{sxy^2} (1-y) \>.
\ee
The usual integrated transversity distribution is given, according to 
some common notations, by:
\be
\Delta_T q(x) \equiv h_{1q}(x) = \int \!\! d^2 \bkt \> \Delta_T q(x, \kt) \>.
\ee

This analysis, performed at ${\cal O}(k_\perp /Q)$, can be further simplified 
adopting a Gaussian and factorized parameterization of the TMDs. In particular 
for the unpolarized parton distribution (TMD-PDFs) and fragmentation  
(TMD-FFs) functions we use:
\bea
 f_{q/p}(x,k_\perp) & = &
 f_{q/p}(x)\;\frac{e^{-{k_\perp^2}/\avk}}{\pi\avk} \label{funp} \\
 D_{h/q}(z,\pp)& =& D_{h/q}(z)\;\frac{e^{-\pp^2/\avp}}{\pi\avp}
 \label{dunp}\;,
\eea
with $\langle k_\perp^2\rangle$ and $\langle p_\perp^2\rangle$ fixed
to the values found in Ref.~\cite{Anselmino:2005nn} by analyzing
unpolarized SIDIS azimuthal dependent data:
\be
\langle k_\perp^2\rangle = 0.25  \; {\rm
GeV}^2 \quad\quad \langle p_\perp^2\rangle  = 0.20 \;{\rm GeV}^2 \,.
\ee
The integrated parton distribution and fragmentation functions,
$f_{q/p}(x)$ and $D_{h/q}(z)$, are available in the literature; in
particular, we use the GRV98LO PDF set~\cite{Gluck:1998xa} and the
DSS fragmentation function set~\cite{deFlorian:2007aj}.

{F}or the transversity distribution, $\Delta_T q(x, k_\perp)$, and
the Collins FF, $\Delta^N\! D_{h/q^\uparrow}(z,\pp)$, we adopt the
following parameterizations~\cite{Anselmino:2007fs}:
 \bea
\Delta_T q(x, k_\perp) &=&\frac{1}{2} \, {\cal N}^{\T}_q(x)\,
[f_{q/p}(x)+\Delta q(x)] 
\> \frac{e^{-{k_\perp^2}/{\avk\T}}}{\pi \avk \T} \label{tr-funct}\\
 \Delta^N \! D_{h/q^\uparrow}(z,\pp) &=& 2 \, {\cal N}^{\C}_{q}(z)\,
D_{h/q}(z) \> h(\pp)\,\frac{e^{-\pp^2/{\avp}}}{\pi \avp}\,,
\label{coll-funct}
\eea
with
\bea
 {\cal N}^{\T}_q(x)= N^{\T}_q
\,x^{\alpha} (1-x)^{\beta} \, \frac{(\alpha + \beta)^{(\alpha
+\beta)}} {\alpha^{\alpha} \beta^{\beta}} \label{NT-transv}
\\
{\cal N}^{\C}_q(z)= N^{\C}_q \, z^{\gamma} (1-z)^{\delta} \,
\frac{(\gamma + \delta)^{(\gamma +\delta)}}
{\gamma^{\gamma} \delta^{\delta}} \label{N(z)-Collins}\label{NC-coll}\\
h(\pp)=\sqrt{2e}\,\frac{p_\perp}{M_{h}}\,e^{-{p_\perp^2}/{M_{h}^2}}
\label{hpcollins}\,,
\eea
and $-1\le N^{\T}_q\le 1$, $-1 \le N^{\C}_q \le 1$. We assume
$\avk \T = \avk$. The combination $[f_{q/p}(x)+\Delta q(x)]$, where 
$\Delta q(x)$ is the helicity distribution, is evolved in $Q^2$ 
according to Ref.~\cite{Vogelsang:1997ak}. Notice that with these 
choices both the transversity and the Collins function automatically 
obey their proper positivity bounds. A different functional form of 
${\cal N}^{\C}_q(z)$ will be explored in Section~\ref{par-poly}.

Using these parameterizations we obtain the following expression for
$A^{\sin (\phi_h+\phi_S)}_{UT}$:
\be
A^{\sin (\phi_h+\phi_S)}_{UT} =
\frac{\displaystyle  \frac{P_T}{M_h}\,\frac{1-y}{s x y^2}\,
\sqrt{2e} \, \frac{\avp ^2 \C}{\avp}
\, \frac{e^{-P_T^2/\avPT \C}}{\avPT ^2 \C} \sum_q e_q^2 \;
 {\cal N}^{\T}_q(x)
\left[f_{q/p}(x)+\Delta q(x) \right]\;
{\cal N}^{\C}_q(z)\,
D_{h/q}(z)}
{ \displaystyle \frac{e^{-P_T^2/\avPT}}{\avPT} \,
\frac{[1+(1-y)^2]}{s x y^2}\,
 \sum_q e_q^2 \, f_{q/p}(x)\; D_{h/q}(z)}\;,
\label{sin-asym-final-1}
\ee
with
\bea
 \avp\! \C= \frac{M_{h}^2 \, \avp}{M_{h}^2 +\avp} &&\hspace*{0.6cm}
 \avPT_{_{\!(C)}} =\avp_{_{\!(C)}} +z^2\avk \,.
\eea
When data or phenomenological information at different $Q^2$ values are 
considered, we take into account, at leading order (LO), the QCD evolution 
of the integrated transversity distribution. For the Collins FF, 
$\Delta ^N\! D_{h/\qup}$, as its scale dependence is unknown, we 
tentatively assume the same $Q^2$ evolution as for the unpolarized FF, 
$D_{h/q}(z)$.

\subsection{$e^+e^-\to h_1 h_2\, X$ processes}

Remarkably, independent information on the Collins functions can be 
obtained in unpolarized $e^+e^-$ processes, by looking at the azimuthal correlations of hadrons produced in opposite jets~\cite{Boer:1997mf}. 
This has been performed by the Belle Collaboration, which have measured 
azimuthal hadron-hadron correlations for inclusive charged pion production, 
$e^+e^-\to \pi \, \pi \, X$~\cite{Abe:2005zx,Seidl:2008xc,Seidl:2012er}. 
This correlation can be interpreted as a direct measure of the Collins 
effect, involving the convolution of two Collins functions.

Two methods have been adopted in the experimental analysis of the Belle 
data. These can be schematically described as (for further details and 
definitions see Refs.~\cite{Boer:1997mf,Anselmino:2007fs,Seidl:2008xc}): 
\\
$i)$ the ``$\cos(\varphi_1 + \varphi_2)$ method'' in the
Collins-Soper frame where the jet thrust axis is used as the $\hat
z$ direction and the $e^+e^-\to q \, \bar q$ scattering defines the
$\widehat{xz}$ plane; $\varphi_1$ and $\varphi_2$ are the azimuthal angles 
of the two hadrons around the thrust axis;
\\
$ii)$ the ``$\cos(2\varphi_0)$ method'', using the Gottfried-Jackson
frame where one of the produced hadrons ($h_2$) identifies the
$\hat z$ direction and the $\widehat{xz}$ plane is determined by the
lepton and the $h_2$ directions. There will then be another relevant
plane, determined by $\hat z$ and the direction of the other
observed hadron $h_1$, at an angle $\varphi_0$ with respect to the
$\widehat{xz}$ plane.

In both cases one integrates over the magnitude of the intrinsic
transverse momenta of the hadrons with respect to the fragmenting
quarks. For the $\cos(\varphi_1 + \varphi_2)$ method the
cross section for the process $e^+e^-\to h_1 \, h_2 \, X$ reads:
 \bea
&&\hspace*{-0.7cm}\frac{d\sigma ^{e^+e^-\to h_1 h_2 X}}
{dz_1\,dz_2\,d\cos\theta\,d(\varphi_1+\varphi_2)}\nonumber\\ [2mm]
&&\hspace*{-0.7cm} =\frac{3\alpha^2}{4s} \, \sum _q e_q^2 \, \Big\{
 (1+\cos^2\theta)\,D_{h_1/q}(z_1)\,D_{h_2/\bar q}(z_2)
\nonumber \\
&&\hspace*{-0.7cm} + \, \Big.\frac{\sin^2\theta}{4} \, 
\cos(\varphi_1\!+\!\varphi_2)\,
\Delta ^N\! D _{h_1/q^\ua}\!(z_1)\, \Delta ^N\! D _{h_2/\bar
q^\ua}\!(z_2)\Big\} \> ,
\label{int-Xs-belle}
 \eea
where $\theta$ is the angle between the lepton direction and the
thrust axis and
 \be
\Delta ^N\! D _{h/q^\ua}(z)\equiv \int d^2\bpp \Delta ^N\!
D_{h/q^\ua} (z,\pp)\,. \label{coll-mom}
 \ee
Integrating over the covered values of $\theta$ and normalizing to the 
corresponding azimuthal averaged unpolarized cross section one has:
\bea
&& R_{12}(z_1,z_2,\varphi_1 + \varphi_2) 
\equiv \frac{1} {\langle d\sigma \rangle} \> \frac{d\sigma
^{e^+e^-\to h_1 h_2 X}}{dz_1\,dz_2\, d(\varphi_1 +
\varphi_2)}
\nonumber \\ [2mm]
&& = 1+\frac{1}{4}\,\frac{
\langle \sin^2\theta \rangle}{\langle 1+\cos^2\theta \rangle}\,
\cos(\varphi_1+\varphi_2) \> \frac{\sum_q
e^2_q \, \Delta ^N\! D_{h_1/q^\ua}(z_1)\,
\Delta ^N\! D_{h_2/\bar q^\ua}(z_2)}{\sum_q e^2_q \, D _{h_1/q}(z_1)\,
D _{h_2/\bar q}(z_2)}\label{A12g}\\
&& \equiv 1 + \frac{1}{4}\,\frac{
\langle \sin^2\theta \rangle}{\langle 1+\cos^2\theta \rangle}\,
\cos(\varphi_1+\varphi_2)\, P(z_1,z_2) \> \cdot \nonumber
\eea

{F}or the $\cos(2\varphi_0)$ method, with the Gaussian ansatz
(\ref{coll-funct}), the analogue of Eq.~(\ref{A12g}) reads
\bea
&& R_0(z_1,z_2,\varphi_0) 
\equiv \frac{1} {\langle d\sigma \rangle} \> \frac{d\sigma
^{e^+e^-\to h_1 h_2 X}}{dz_1\,dz_2\, d\varphi_0}
\nonumber\\ [2mm]
&& = 1+\frac{1}{\pi}\,\frac{z_1\,z_2}{z_1^2+z_2^2}\,
\frac{\langle\sin^2\theta_2\rangle}{\langle 1+\cos^2\theta_2 \rangle}
\,\cos(2\varphi_0) \> \frac{\sum_q 
e^2_q \, \Delta ^N\! D _{h_1/q^\ua}(z_1)\,
\Delta ^N\! D  _{h_2/\bar q^\ua}(z_2)}{\sum_q e^2_q D _{h_1/q}(z_1)\,
D _{h_2/\bar q}(z_2)} \label{A0g} \\
&& \equiv 1+\frac{1}{\pi}\,\frac{z_1\,z_2}{z_1^2+z_2^2}\,
\frac{\langle\sin^2\theta_2\rangle}{\langle 1+\cos^2\theta_2 \rangle}
\,\cos(2\varphi_0) \, P(z_1,z_2) \>, \nonumber
\eea
where $\theta_2$ is now the angle between the lepton and the $h_2$
hadron directions. 

In both cases, Eqs. (\ref{A12g}) and (\ref{A0g}), the value of 
\be
\frac{\langle \sin^2\theta \rangle}
{\langle 1+\cos^2\theta \rangle} \equiv C(\theta) 
\ee
can be found in the experimental data (see Tables IV and V of 
Ref.~\cite{Seidl:2008xc}). 

To eliminate false asymmetries, the Belle Collaboration consider the 
ratio of unlike-sign ($\pi^+\pi^-$ + $\pi^-\pi^+$)
to like-sign ($\pi^+\pi^+$ + $\pi^-\pi^-$) or charged 
($\pi^+\pi^+$ + $\pi^+\pi^- + \pi^-\pi^+$ + $\pi^-\pi^-$)
pion pair production, denoted respectively with indices $U$, $L$ and $C$. 
For example, in the case of unlike- to like-pair production, one has 
\bea
\frac{R_{12}^U}{R_{12}^L} =
\frac{\displaystyle 1 + \frac{1}{4} \, C(\theta) \,
\cos(\varphi_1+\varphi_2)\, P_U}
{\displaystyle 1 + \frac{1}{4} \, C(\theta) \,
\cos(\varphi_1+\varphi_2)\, P_L} \label{R12} 
&\simeq& 1 + \frac{1}{4}\, C(\theta)
\cos(\varphi_1+\varphi_2)\, (P_U - P_L) \\ 
&\equiv& 1 + \cos(\varphi_1+\varphi_2)\, A_{12}^{UL} \label{A12}
\eea  
and
\bea
\frac{R_{0}^U}{R_{0}^L} =
\frac{\displaystyle
1 + \frac{1}{\pi} \, \frac{z_1\,z_2}{z_1^2+z_2^2} \, C(\theta) \,
\cos(2\varphi_0)\, P_U}
{\displaystyle
1 + \frac{1}{\pi} \, \frac{z_1\,z_2}{z_1^2+z_2^2} \, C(\theta) \,
\cos(2\varphi_0)\, P_L} \label{R0} 
&\simeq& 1 + \frac{1}{\pi}\, \frac{z_1\,z_2}{z_1^2+z_2^2} \, C(\theta)
\cos(2\varphi_0)\, (P_U - P_L) \\ 
&\equiv& 1 + \cos(2\varphi_0)\, A_{0}^{UL} \label{A0}
\eea  
and similarly for $R^U_{12}/R^C_{12}$ and $R^U_{0}/R^C_{0}$.
Explicitely, one has:
\bea
&&
P_U= \frac{\sum_q e^2_q \;
    [\Delta ^N D _{\pi^+/q^\ua}(z_1)\,
     \Delta ^N D _{\pi^-/\bar q^\ua}(z_2) +
     \Delta ^N D _{\pi^-/q^\ua}(z_1)\,
     \Delta ^N D _{\pi^+/\bar q^\ua}(z_2)]}
{\sum_q e^2_q \;[D _{\pi^+/q}(z_1)\,D _{\pi^-/\bar q}(z_2) +
                 D _{\pi^-/q}(z_1)\,D _{\pi^+/\bar q}(z_2)]} \equiv 
     \frac{(P_U)_N}{(P_U)_D} \label{PU}
\\
&&
P_L= \frac{\sum_q e^2_q \;
    [\Delta ^N D _{\pi^+/q^\ua}(z_1)\,
     \Delta ^N D _{\pi^+/\bar q^\ua}(z_2) +
     \Delta ^N D _{\pi^-/q^\ua}(z_1)\,
     \Delta ^N D _{\pi^-/\bar q^\ua}(z_2)]}
{\sum_q e^2_q \;[D _{\pi^+/q}(z_1)\,D _{\pi^+/\bar q}(z_2) +
                 D _{\pi^-/q}(z_1)\,D _{\pi^-/\bar q}(z_2)]} \equiv
     \frac{(P_L)_N}{(P_L)_D}
\\
&&
P_C = \frac{(P_U)_N + (P_L)_N}{(P_U)_D + (P_L)_D} \\
&&
A_{12}^{UL,C}(z_1,z_2)=\frac{1}{4}\frac{\langle \sin^2\theta \rangle}
{\langle 1+\cos^2\theta \rangle}\,(P_U - P_{L,C} )\label{A12ulc} \\
&&
A_{0}^{UL,C}(z_1,z_2)=\frac{1}{\pi} \,
\frac{z_1\,z_2}{z_1^2+z_2^2} \frac{\langle \sin^2\theta_2 \rangle}
{\langle 1+\cos^2\theta_2 \rangle}\,(P_U - P_{L,C} ) \>.\label{A0ulc}
\eea
For fitting purposes, it is convenient to introduce favoured and 
disfavoured fragmentation functions, assuming in Eq.~(\ref{coll-funct}):
\bea
&& \frac{\Delta^ND_{\pi^+/u^\ua,\bar d^\ua}(z, \pp)}{D_{\pi^+/u, \bar d}(z)}
=  \frac{\Delta^ND_{\pi^-/d^\ua,\bar u^\ua}(z, \pp)}{D_{\pi^-/d, \bar u}(z)}
=  2 \, {\cal N}^{\C}_{\rm fav}(z)\> h(\pp) \,
   \frac{e^{-\pp^2/{\avp}}}{\pi \avp} \label{fav} \\
&& \frac{\Delta^ND_{\pi^+/d^\ua,\bar u^\ua}(z, \pp)}{D_{\pi^+/d, \bar u}(z)}
=  \frac{\Delta^ND_{\pi^-/u^\ua,\bar d^\ua}(z, \pp)}{D_{\pi^-/u, \bar d}(z)}
=  \frac{\Delta^ND_{\pi^\pm/s^\ua,\bar s^\ua}(z, \pp)}{D_{\pi^\pm/s, \bar s}(z)}
=  2 \, {\cal N}^{\C}_{\rm dis}(z)\> h(\pp) \,
   \frac{e^{-\pp^2/{\avp}}}{\pi \avp} \>, \label{unf}
\eea
with the corresponding relations for the integrated Collins functions,
Eq.~(\ref{coll-mom}), and with ${\cal N}^{\C}_{\rm fav, dis}(z)$ as 
given in Eq.~(\ref{N(z)-Collins}) with $N^{\C}_q = N^{\C}_{\rm fav, dis}$.

We can now perform a best fit of the data from HERMES and COMPASS on
$A_{UT}^{\sin(\phi_h + \phi_S)}$ and of the data, from the Belle 
Collaboration, on $A_{12}^{UL,C}$ and $A_{0}^{UL,C}$. Their expressions,
Eqs.~(\ref{sin-asym-final-1}) and (\ref{PU})--(\ref{unf}), contain the transversity and the Collins functions, parameterised as in 
Eqs.~(\ref{tr-funct})--(\ref{hpcollins}). They depend on the free 
parameters $\alpha, \beta, \gamma, \delta, N^{\T}_q, N^{\C}_q$ and $M_h$. 
Following Ref.~\cite{Anselmino:2007fs} we assume the exponents $\alpha, \beta$ 
and the mass scale $M_h$ to be flavour independent and consider the transversity 
distributions only for $u$ and $d$ quarks (with the two free parameters 
$N^{\T}_u$ and $N^{\T}_d$). The favoured and disfavoured Collins functions 
are fixed, in addition to the flavour independent exponents $\gamma$ and 
$\delta$, by $N^{\C}_{\rm fav}$ and  $N^{\C}_{\rm dis}$. This makes a total 
of 9 parameters, to be fixed with a best fit procedure.   
Notice that while in the present analysis we can safely neglect 
any flavour dependence of the parameter $\beta$ (which is anyway hardly 
constrained by the SIDIS data), this issue could play a significant role 
in other studies, like those discussed in Ref.~\cite{Anselmino:2012rq}.

\section{Best fits, results and parameterisations}
\subsection{Standard parameterisation}

We start by repeating the same fitting procedure as in 
Refs.~\cite{Anselmino:2007fs, Anselmino:2008jk}, using the same 
``standard" parameterisation, Eqs.~(\ref{funp})--(\ref{hpcollins}), 
with the difference that now we include all the most recent SIDIS data 
from COMPASS~\cite{Martin:2013eja} and HERMES~\cite{Airapetian:2010ds} Collaborations, and the corrected Belle data~\cite{Seidl:2012er} on 
$A_{12}^{UL}$ and $A_{12}^{UC}$. Notice, in particular, that the 
$A_{12}^{UC}$ data are included in our fits for the first time here.
In fact, a previous inconsistency between $A_{12}^{UL}$ and $A_{12}^{UC}$
data, present in the first Belle results~\cite{Abe:2005zx},
has been removed in Ref.~\cite{Seidl:2012er}.

\begin{figure}[t]
\begin{center}
\includegraphics[width=0.4\textwidth, angle=-90]{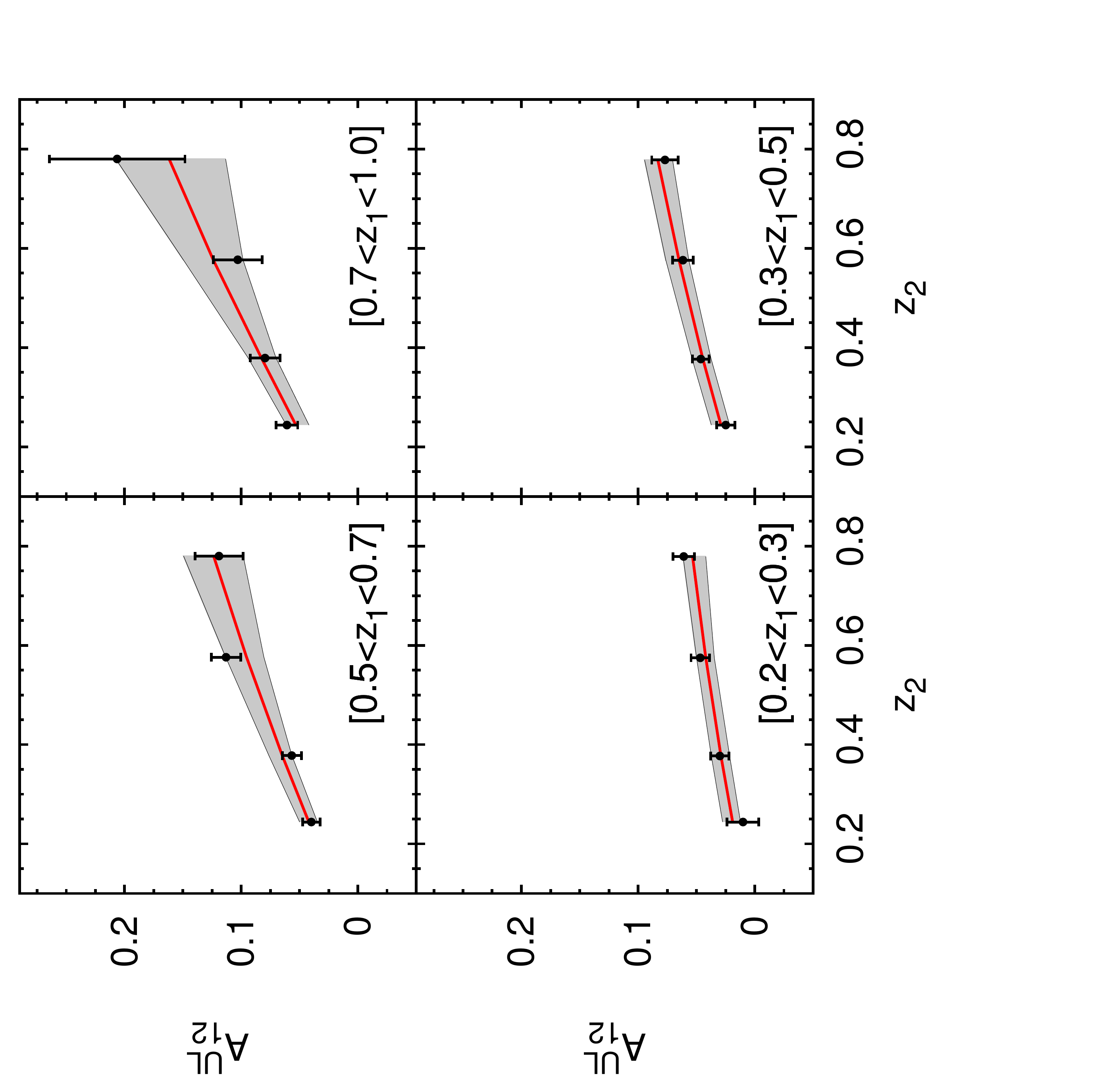}
\hspace*{.3cm}
\includegraphics[width=0.4\textwidth, angle=-90]{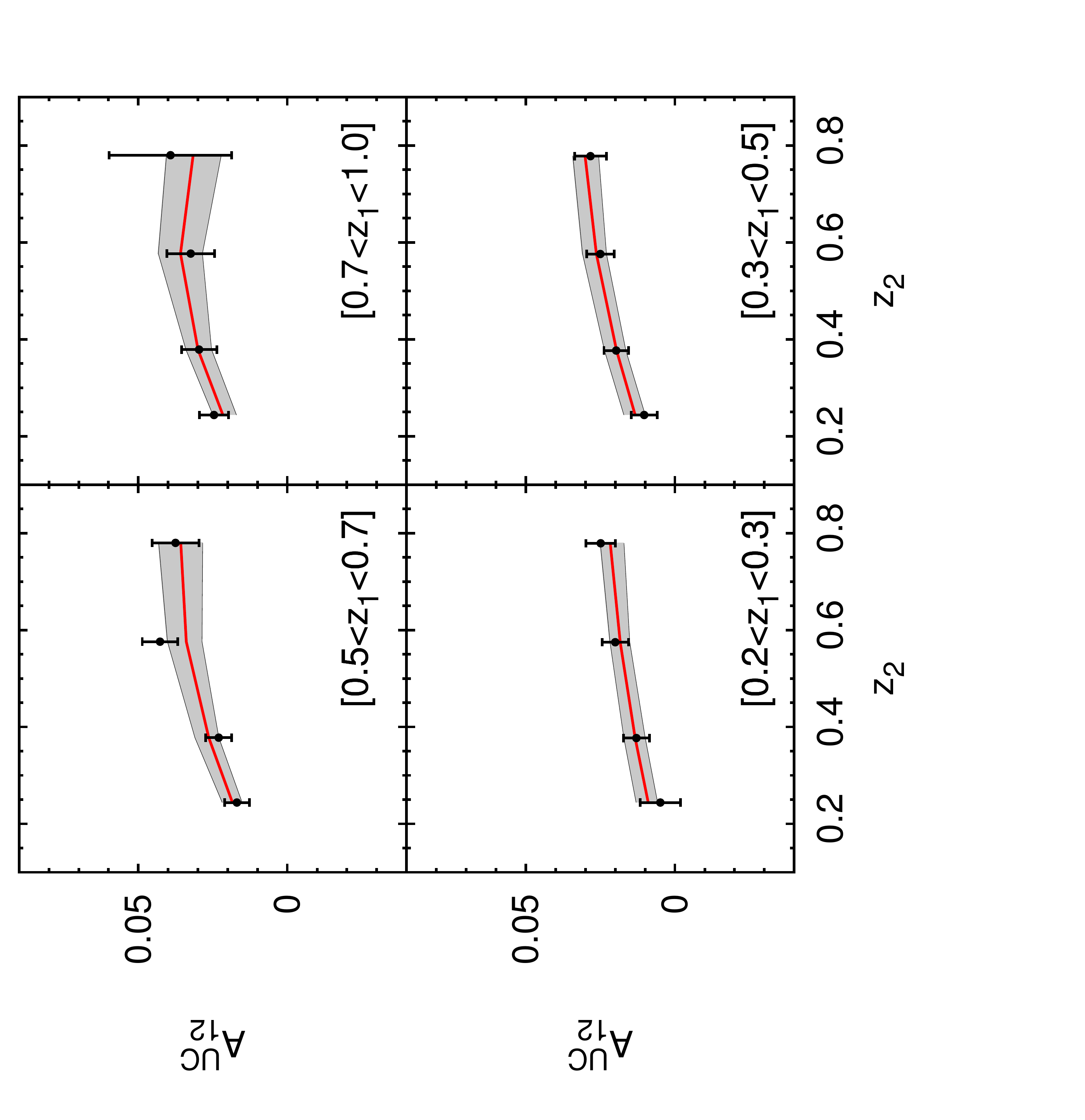}
\\
\includegraphics[width=0.4\textwidth, angle=-90]{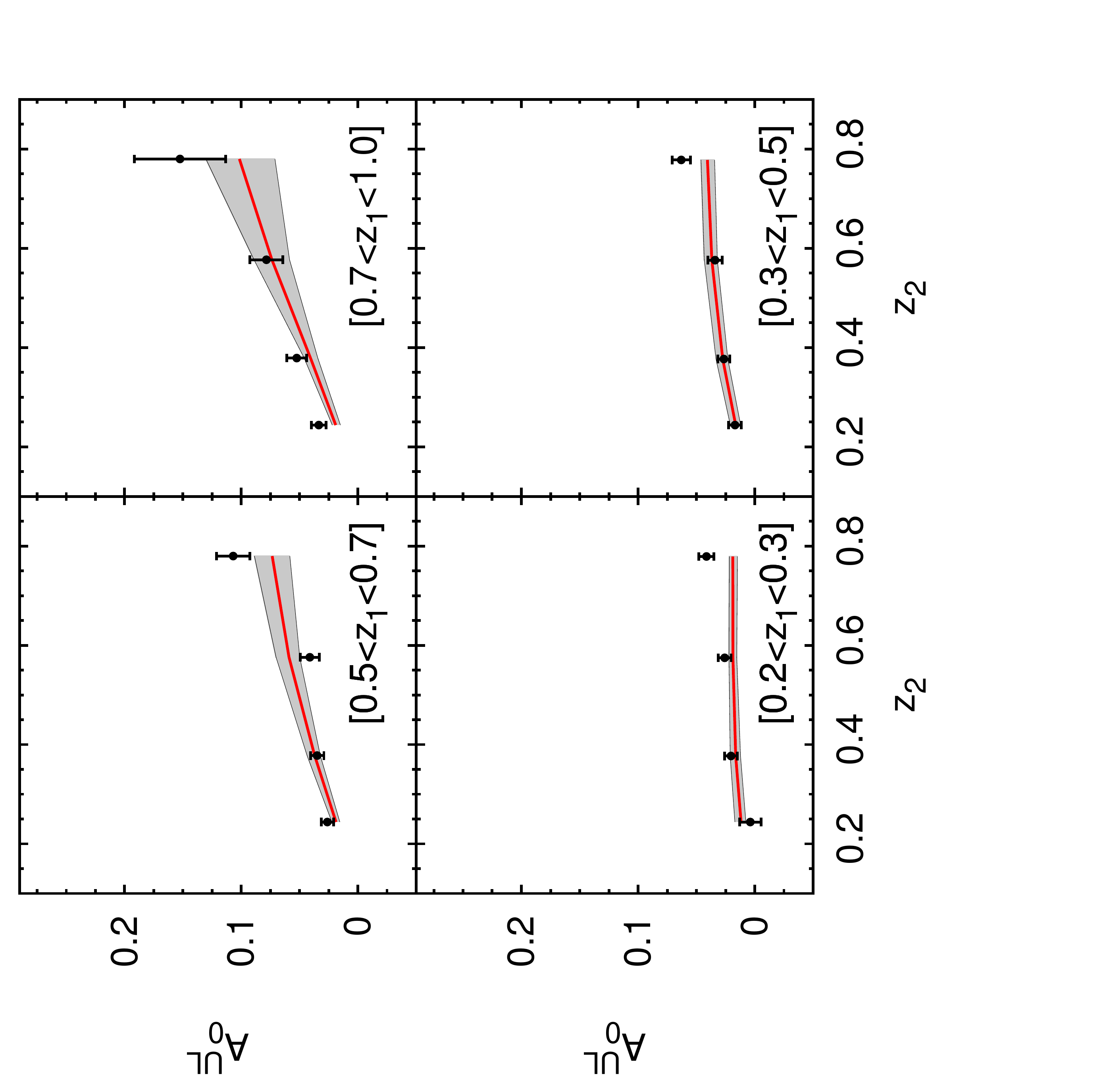}
\hspace*{.3cm}
\includegraphics[width=0.4\textwidth, angle=-90]{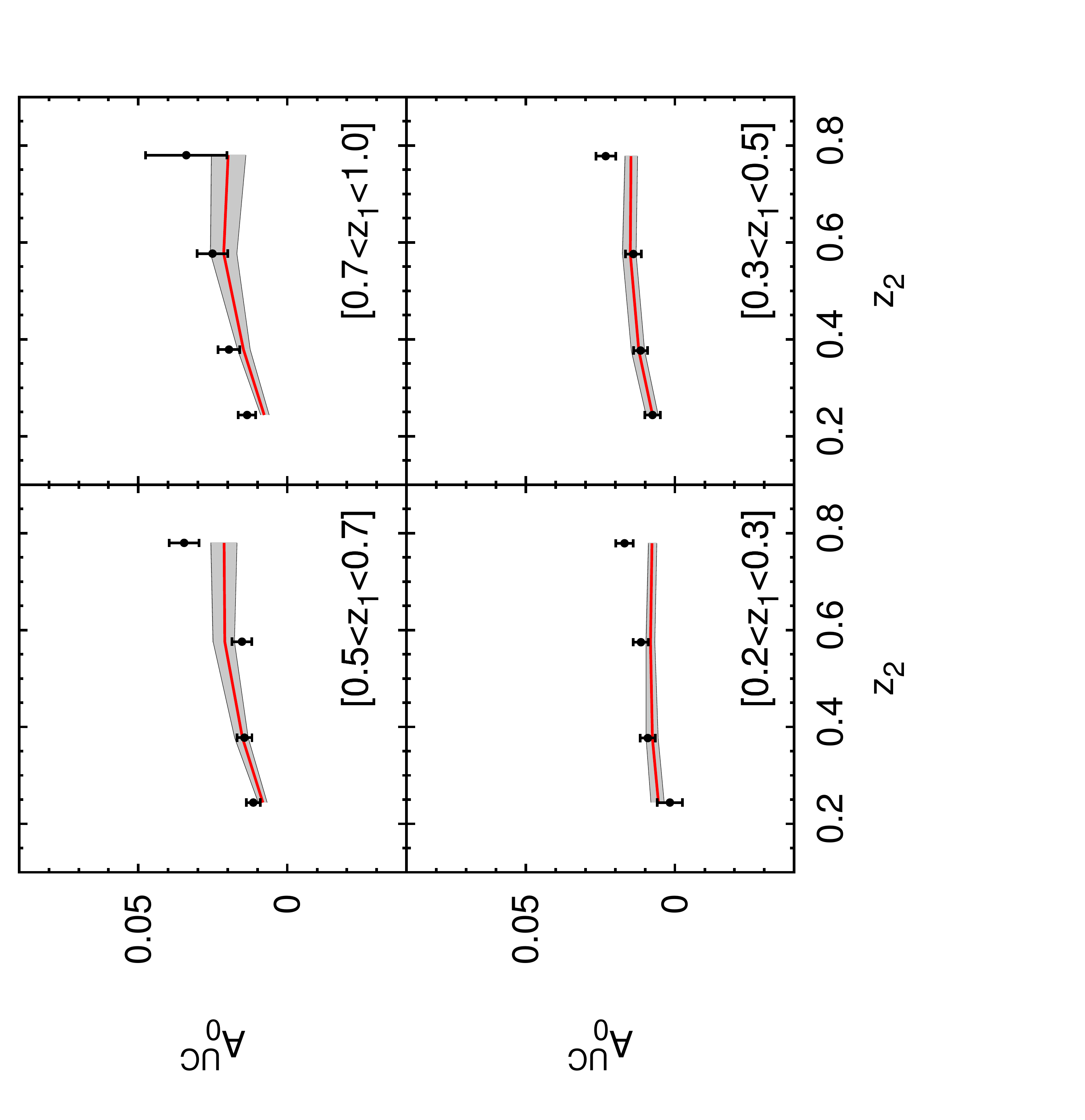}
\vskip -18pt 
\caption{\label{fig:Belle-standard-A12fit}
The experimental data on $A_{12}^{UL}$, $A_{12}^{UC}$ (upper plots) 
and $A_{0}^{UL}$ and $A_{0}^{UC}$ (lower plots), as measured 
by the Belle Collaboration~\cite{Seidl:2012er} in unpolarized 
$e^+e^- \to h_1 \, h_2 \, X$ processes, are compared to 
the curves obtained from our global fit. The solid lines correspond 
to the parameters given in Table~\ref{fitpar}, obtained by fitting the 
SIDIS and the $A_{12}$ asymmetries with the standard parameterisation; 
the shaded areas correspond to the statistical uncertainty on the 
parameters, as explained in the text and in Ref.~\cite{Anselmino:2008sga}.
Notice that the $A_{0}^{UL}$ and $A_{0}^{UC}$ data are not included in 
the fit and our curves, with the corresponding uncertainties, are 
simply computed using the parameters of Table~\ref{fitpar}. 
}
\end{center}
\end{figure}

\begin{figure}[t]
\begin{center}
\includegraphics[width=0.46\textwidth, angle=-90]{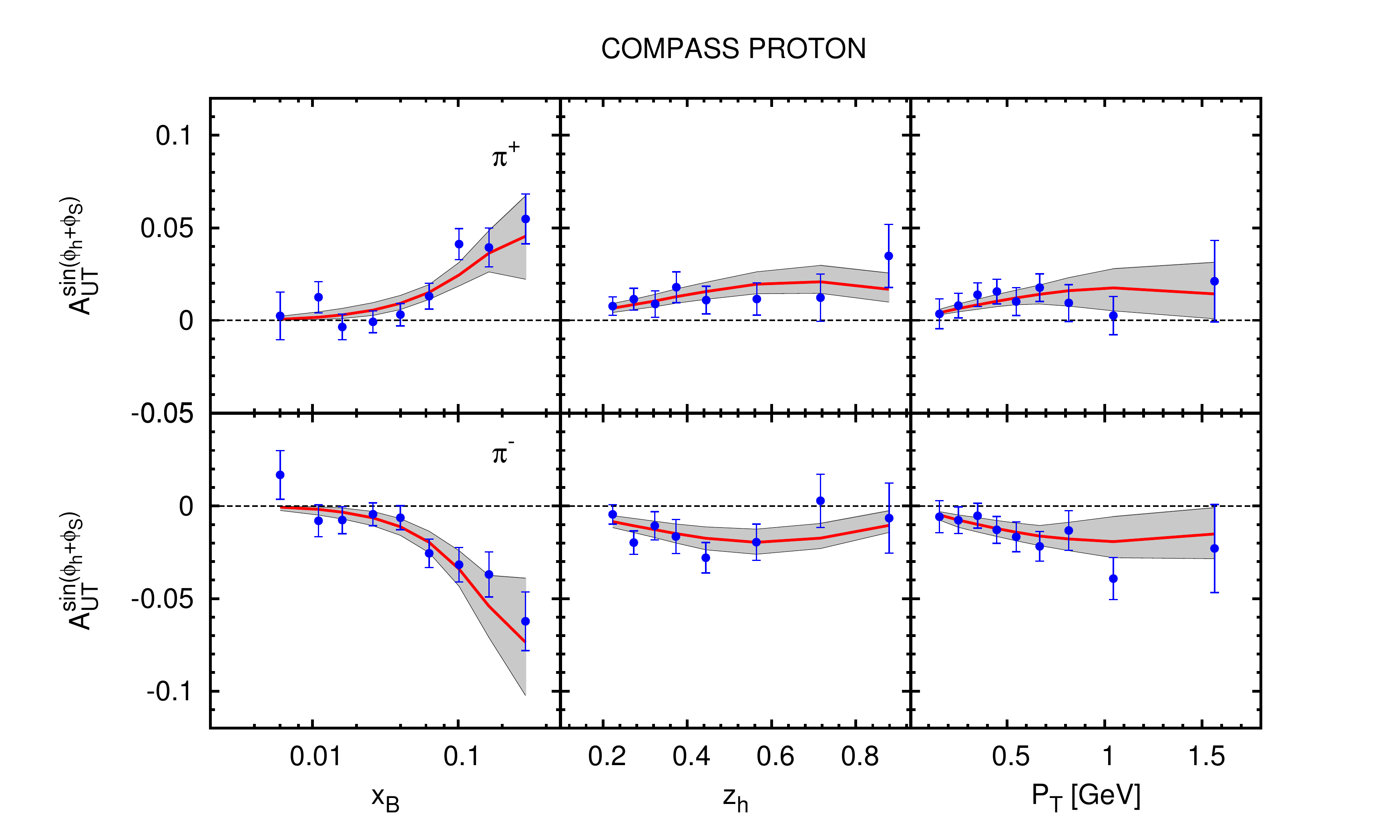}
\hspace*{.3cm}
\includegraphics[width=0.46\textwidth, angle=-90]{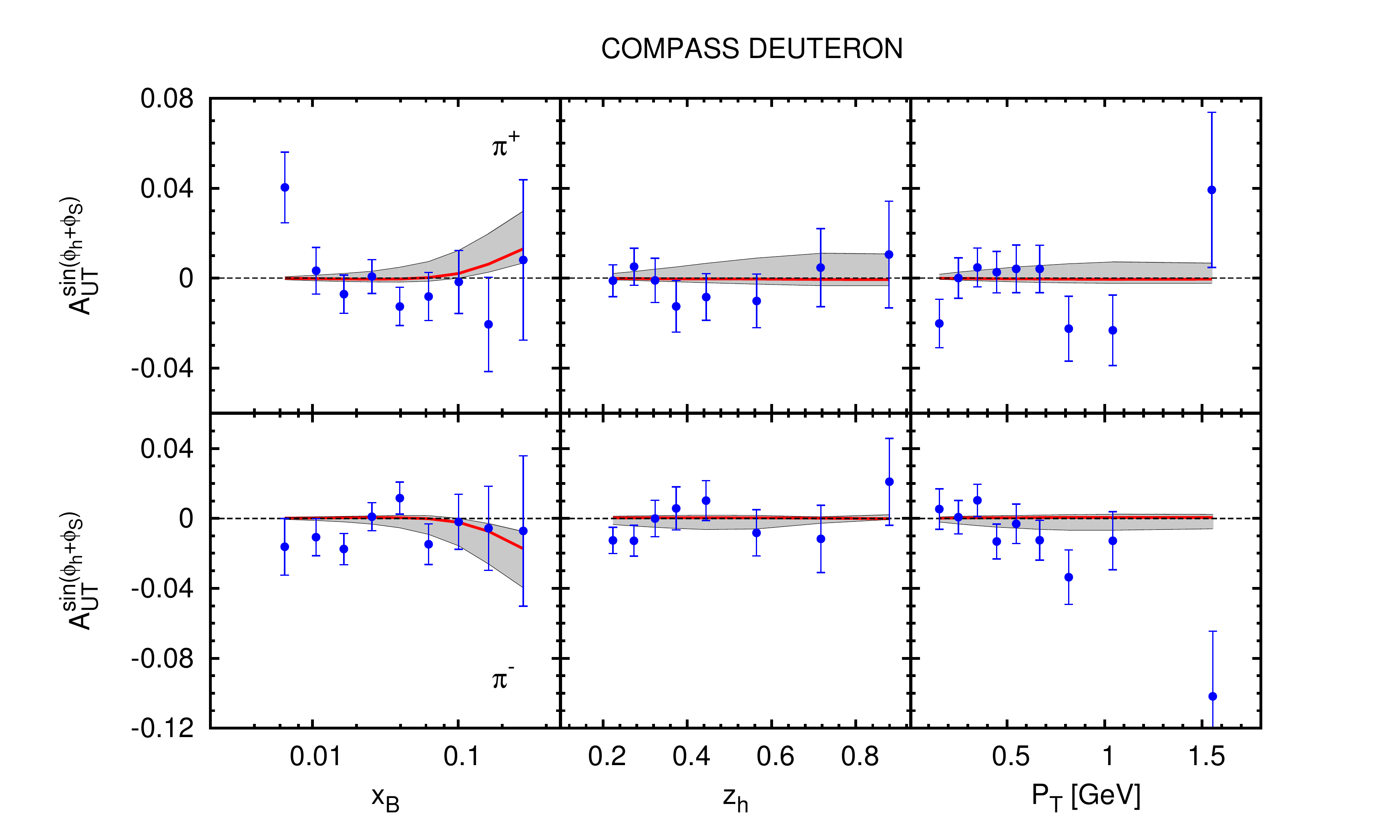}
\caption{\label{fig:compass-p-d}
The experimental data on the SIDIS azimuthal moment $A_{UT}^{\sin(\phi_h +
\phi_S)}$ as measured by the COMPASS Collaboration~\cite{Martin:2013eja} 
on proton (upper plots) and deuteron (lower plots) targets, are compared 
to the curves obtained from our global fit.
The solid lines correspond to the parameters given in Table~\ref{fitpar}, obtained by fitting the SIDIS and the $A_{12}$ asymmetries with standard 
parameterisation; the shaded areas correspond to the statistical 
uncertainty on the parameters, as explained in the text and in 
Ref.~\cite{Anselmino:2008sga}.
}
\end{center}
\end{figure}

\begin{figure}[t]
\begin{center}
\includegraphics[width=0.46\textwidth, angle=-90]{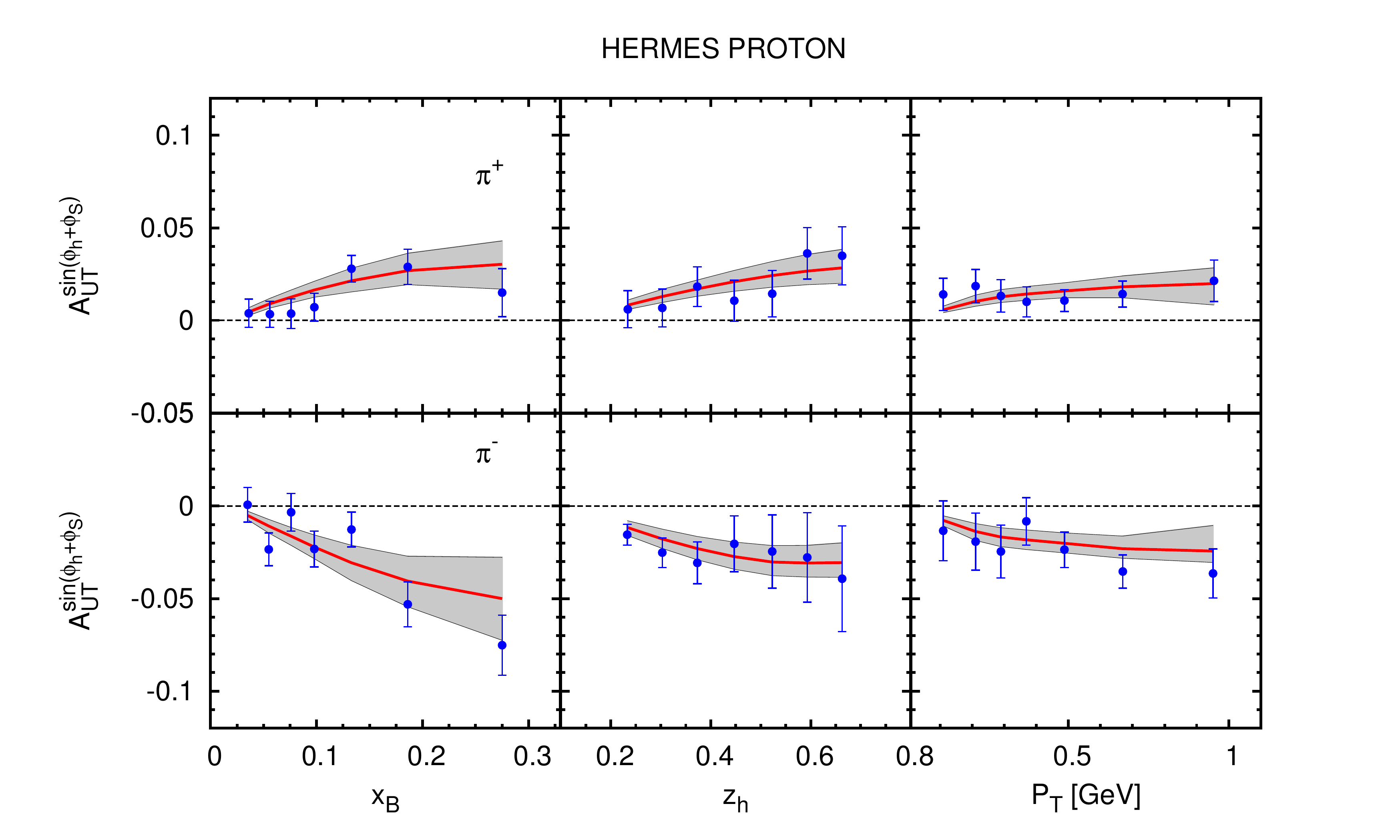}
\caption{\label{fig:hermes-pi}
The experimental data on the SIDIS azimuthal moment $A_{UT}^{\sin(\phi_h +
\phi_S)}$ as measured by the HERMES Collaboration~\cite{Airapetian:2010ds}, 
are compared to the curves obtained from our global fit. The solid lines 
correspond to the parameters given in Table~\ref{fitpar}, obtained by 
fitting the SIDIS and the $A_{12}$ asymmetries with standard 
parameterisation; the shaded areas correspond to the statistical 
uncertainty on the parameters, as explained in the text and in 
Ref.~\cite{Anselmino:2008sga}.
}
\end{center}
\end{figure}

The results we obtain are remarkably good, with a total $\chi^2_{\rm d.o.f}$ 
of $0.80$, as reported in the first line of Table~\ref{chisq}, 
and the values of the resulting parameters, given in Table~\ref{fitpar}, 
are consistent with those found in our previous extractions. Our best 
fits are shown in Fig.~\ref{fig:Belle-standard-A12fit} (upper plots), for 
the Belle $A_{12}$ data, in Fig.~\ref{fig:compass-p-d} for the SIDIS 
COMPASS data and in Fig.~\ref{fig:hermes-pi} for the HERMES results. 

\begin{table}[t]
\caption{Summary of the $\chi ^2$ values obtained in our fits.
The columns, from left to right give the $\chi ^2$ per degree of freedom, 
the total $\chi ^2$, and the separate contributions to the total $\chi ^2$ 
of the data from SIDIS, $A_{12}^{UL}$, $A_{12}^{UC}$, $A_{0}^{UL}$ and 
$A_{0}^{UC}$. ``NO FIT" means that the $\chi ^2$ for that set of data 
does not refer to a best fit, but to the computation of the corresponding 
quantity using the best fit parameters fixed by the other data. 
The four lines show the results for the two choices of parameterisation of 
the $z$ dependence of the Collins functions (standard and polynomial) 
and for the two independent sets of data fitted (SIDIS, $A_{12}^{UL}$,
$A_{12}^{UC}$ and SIDIS, $A_{0}^{UL}$, $A_{0}^{UC}$).
\label{chisq}}
\vskip 18pt
\renewcommand{\tabcolsep}{0.4pc} 
\renewcommand{\arraystretch}{1.2} 
\begin{tabular}{@{}|c|c|c|c|c|c|c|}
 \hline
 ~                        & ~             & ~ & ~               & ~               & ~ & ~              \\
 ~                        & FIT DATA      & SIDIS &~~$A_{12}^{UL}$~~&~~$A_{12}^{UC}$~~&~~$A_{0}^{UL}$~~&~~$A_{0}^{UC}$~~\\
 ~                        & 178  points   & 146  points & 16 points       & 16 points       & 16 points & 16  points      \\
 \hline
 Standard                 & ~             & ~ & ~               & ~               & ~ & ~              \\
 Parameterization         & $\chi ^2_{\rm tot} = 135$  & $\chi ^2 = 123$        & $\chi ^2 =7$               & $\chi ^2 =5$           & $\chi ^2 = 44$           & $\chi ^2 = 39$             \\
 $\chi ^2_{\rm d.o.f} = 0.80$ & ~             & ~ & ~               & ~               & NO FIT & NO FIT         \\
 \hline
 Standard                 & ~             & ~ & ~               & ~               & ~            & ~                \\
 Parameterization         & $\chi ^2_{\rm tot} = 190$ & $\chi ^2 = 125$           & $\chi ^2 = 20$            & $\chi ^2 = 12$          & $\chi ^2 = 35$           & $\chi ^2 = 30$               \\
 $\chi ^2_{\rm d.o.f} = 1.12$ & ~             & ~ & NO FIT          & NO FIT          & ~            & ~                \\
 \hline
 \hline
 Polynomial               & ~             & ~ & ~               & ~               & ~            & ~                \\
 Parameterization         & $\chi ^2_{\rm. tot} = 136$ & $\chi ^2 = 123$           & $\chi ^2 = 8$            & $\chi ^2 = 5$            & $\chi ^2 = 45$           & $\chi ^2 = 39$               \\
 $\chi ^2_{\rm d.o.f} = 0.81$ & ~             & ~ & ~               & ~               & NO FIT       & NO FIT           \\
 \hline
 Polynomial               & ~             & ~ & ~               & ~               & ~            & ~                \\
 Parameterization         & $\chi ^2_{\rm tot} = 171$ & $\chi ^2 = 141$           & $\chi ^2 = 44$           & $\chi ^2 = 27$           & $\chi ^2 = 15$           & $\chi ^2 = 15$               \\
 $\chi ^2_{\rm d.o.f} = 1.01$ & ~             & ~ & NO FIT          & NO FIT          & ~            & ~                \\
 \hline
\end{tabular}
\end{table} 
\begin{table}[b]
\caption{
Best values of the 9 free parameters fixing the $u$ and $d$ quark
transversity distribution functions and the favoured and
disfavoured Collins fragmentation functions, as obtained by fitting 
simultaneously SIDIS data on the Collins asymmetry and Belle data on 
$A_{12}^{UL}$ and $A_{12}^{UC}$. The transversity distributions are parameterised according to Eqs.~(\ref{tr-funct}), (\ref{NT-transv}) and the Collins 
fragmentation functions according to the standard parameterisation, 
Eqs.~(\ref{coll-funct}), (\ref{NC-coll}) and (\ref{hpcollins}). We obtain a 
total $\chi^2/{\rm d.o.f.} = 0.80$. The statistical errors quoted for each 
parameter correspond to the shaded uncertainty areas in 
Figs.~\ref{fig:Belle-standard-A12fit}--\ref{fig:hermes-pi},
as explained in the text and in the Appendix of Ref.~\cite{Anselmino:2008sga}.
\label{fitpar}}
\vskip 18pt
\renewcommand{\tabcolsep}{0.4pc} 
\renewcommand{\arraystretch}{1.2} 
\begin{tabular}{@{}ll}
 \hline
 $N_{u}^{\T}$ = $0.46^{+0.20}_{-0.14}$ & $N_{d}^{\T}$ = $ -1.00^{+1.17}_{-0.00}$ \\
 $\alpha$ =  $1.11^{+0.89}_{-0.66}$ & $\beta$  = $3.64^{+5.80}_{-3.37}$ \\
 \hline
 $N_{\rm fav}^{\C}$  = $0.49^{+0.20}_{-0.18}$ & $N_{\rm dis}^{\C}$  = 
 $-1.00^{+0.38}_{-0.00}$ \\
 $\gamma$  = $1.06^{+0.45}_{-0.32}$  & $\delta$   = $0.07^{+0.42}_{-0.07}$    \\
 $M^2_h = 1.50^{+2.00}_{-1.12}$ GeV$^2$ & \\
 \hline
\end{tabular}
\end{table}

We have not inserted the $A_{0}$ Belle data in our global analysis 
as they are strongly correlated with the $A_{12}$ results, being a 
different analysis of the same experimental events. However,  
using the extracted parameters we can compute the $A_{0}^{UL}$ and 
$A_{0}^{UC}$ azimuthal asymmetries, in good qualitative agreement with 
the Belle measurements, although the corresponding $\chi ^2$ values are 
rather large, as shown in Table~\ref{chisq}. These results are 
presented in Fig.~\ref{fig:Belle-standard-A12fit} (lower plots).

The shaded uncertainty bands are computed according to the procedure 
explained in the Appendix of Ref.~\cite{Anselmino:2008sga}. We have 
allowed the set of best fit parameters to vary in such a way that 
the corresponding new curves have a total $\chi^2$ which differs from 
the best fit $\chi^2$ by less than a certain amount $\Delta\chi^2$. 
All these (1500) new curves lie inside the shaded area. The chosen value 
of $\Delta\chi^2 = 17.21$ is such that the probability to find the ``true" 
result inside the shaded band is 95.45\%. 

We have also performed a global fit based on the SIDIS and $A_{0}$ Belle
data, and then computed the $A_{12}$ values. We do not show the best fit 
plots, which are not very informative, but the quality of the results 
can be judged from the second line of Table~\ref{chisq}, which shows that 
although this time $A_{0}^{UL}$ and $A_{0}^{UC}$ are actually 
fitted, their corresponding $\chi ^2$ values remain large. This has 
induced us to explore a different functional shape for the 
parameterisation of ${\cal N}^{\C}_q(z)$, Eq.~(\ref{N(z)-Collins}),
which will be discussed in the next Subsection.   


The difference between $A_{12}$ and $A_{0}$ is a delicate issue, that 
deserves some further comments. On the experimental side, the hadronic-plane 
method used for the extraction of $A_{0}$ implies a simple analysis 
of the raw data, as it requires the sole reconstruction of the 
tracks of the two detected hadrons; therefore it leads to very clean 
data points, with remarkably small error bars.
On the contrary, the thrust-axis method is much more involved as it 
requires the reconstruction of the original direction of the $q$ and 
$\bar q$ which fragment into the observed hadrons; this makes the 
measurement of the $A_{12}$ asymmetry experimentally more challenging, 
and leads to data points whith larger uncertainties. 

On the theoretical side, the situation is just the opposite: as the 
thrust-axis method assumes a perfect knowledge of the $q$ and $\bar q$ 
directions, the asymmetry can be reconstructed by a straightforward 
integration over the two intrinsic transverse momenta $p_{\perp 1}$ 
and $p_{\perp 2}$, transforming the convolution of two Collins functions  
into the much simpler product of two Collins 
moments~\cite{Anselmino:2007fs}, Eqs.~(\ref{coll-mom}) and (\ref{A12g}). 
Instead, the phenomenological partonic expression of $A_{0}$ requires 
more assumptions and approximations and is, on the theoretical side, 
less clean. 

One should also add that most of the large $\chi^2$ values found when 
computing $A_0$ from the parameters of a best fit involving SIDIS
and $A_{12}$ data (or vice-versa) originate from the experimental
points at large values of $z_1$ or $z_2$ or both (see, for example 
the last points on the left lower panel in 
Fig.~\ref{fig:Belle-standard-A12fit}). Large values of $z$ bring us
near the exclusive process limit, where our factorized inclusive
approach cannot hold anymore.        
 
\subsection{Polynomial parameterisation}\label{par-poly}

In an attempt to fit equally well $A_{12}$ and $A_{0}$
(keeping in mind, however, the comments at the end of the previous 
Subsection) we have explored a possible new parameterisation of the 
$z$ dependence of the Collins function. We notice that data on 
$A_{0}(z)$ seem to favour an increase at large $z$ values, rather 
then a decrease, which is implicitly forced by a behaviour of the kind 
given in Eqs.~(\ref{coll-funct}) and (\ref{N(z)-Collins}) (at least
with positive $\delta$ values).      

In addition, an increasing trend of $A_0(z)$ and $A_{12}(z)$ seems 
to be confirmed by very interesting preliminary results of the BABAR 
Collaboration, which have performed an independent new analysis of 
$e^+ e^- \to h_1 \, h_2 \, X$ data~\cite{Garzia:2012za}, analogous to 
that of Belle.

This suggests that a different parameterisation of the $z$ dependence 
of favoured and disfavoured Collins functions could turn out to be more 
convenient. Then, we try an alternative polynomial parameterisation which 
allows more flexibility on the behaviour of ${\cal N}^{\C}_q(z)$ at 
large $z$:
\be
{\cal N}^{\C}_q(z)= N^{\C}_q\,z\,[(1-a - b) + a\,z + b\,z^2] 
\label{coll-2} \,,
\ee
with the subfix $q={\rm fav, dis}$, and $-1\le N^{\C}_q \le1$; $a$ and 
$b$ are flavour independent so that the total number of parameters for
the Collins functions (in addition to $M_h$) remains 4. Such a choice
fixes the term ${\cal N}^{\C}_q(z)$ to be equal to $0$ at $z=0$ and 
not larger than $1$ at $z=1$. Notice that we do not automatically impose, 
as in Eq.~(\ref{N(z)-Collins}), the condition $|{\cal N}^{\C}_q(z)| \leq 1$;
however, we have explicitly checked that the best fit results and all 
the sets of parameters corresponding to curves inside the shaded 
uncertainty bands satisfy that condition.   

\begin{figure}[t]
\begin{center}
\includegraphics[width=0.35\textwidth, angle=-90]{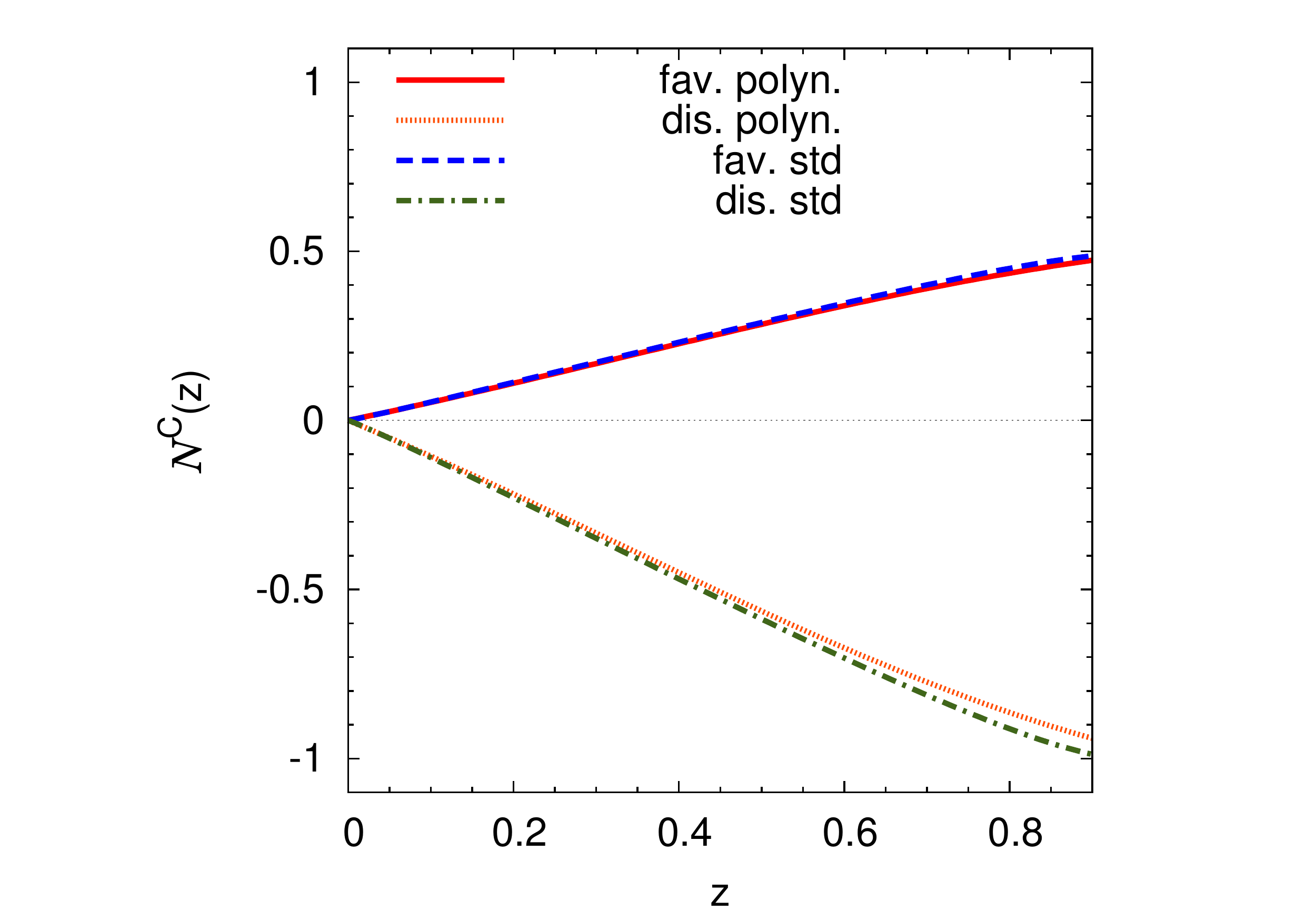}
\hspace*{-.5cm}
\includegraphics[width=0.35\textwidth, angle=-90]{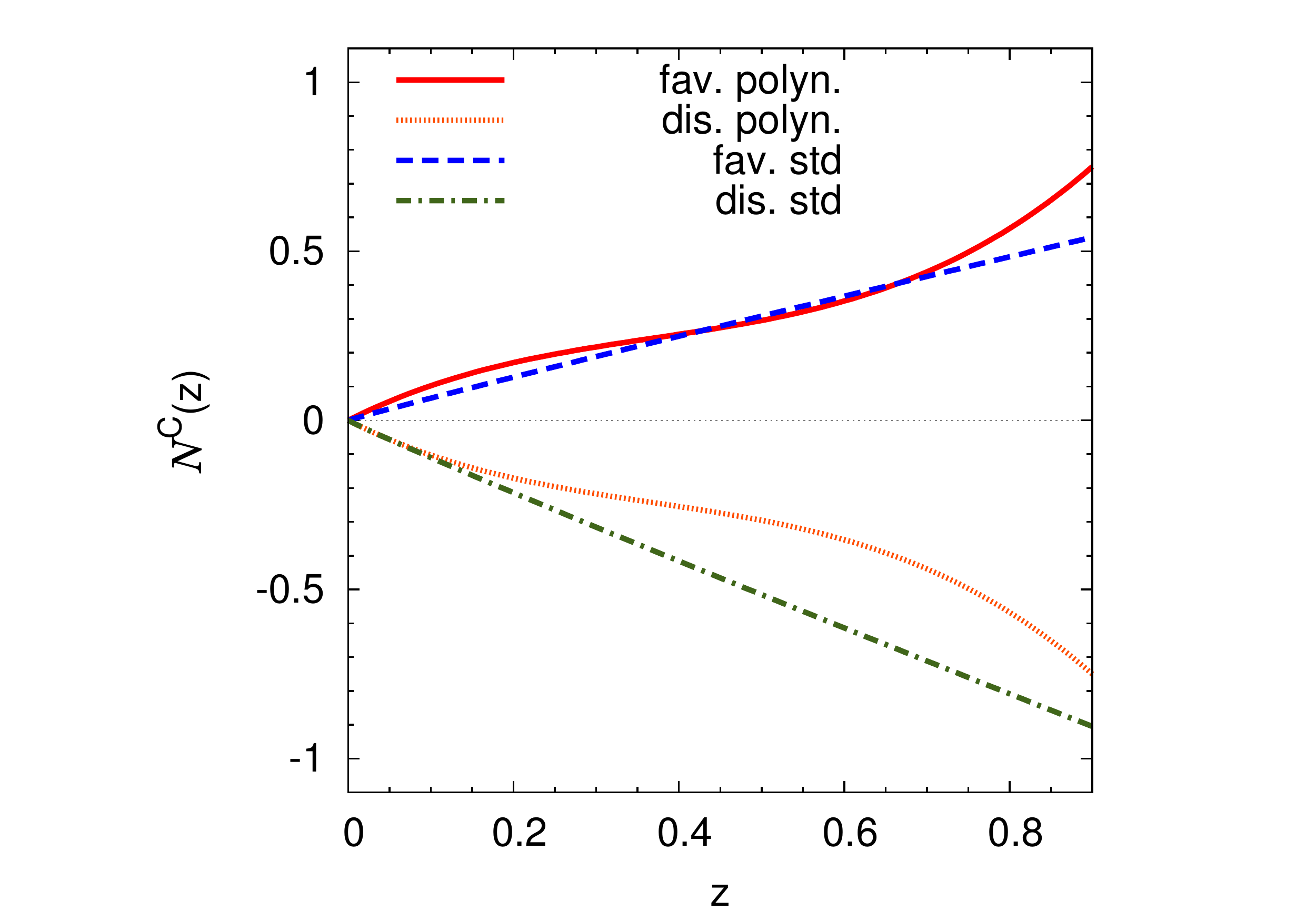}
\caption{\label{fig:poly}
Plots of the functions ${\cal N}^{\C}_{\rm fav}(z)$ and 
${\cal N}^{\C}_{\rm dis}(z)$ for the favoured and disfavoured Collins 
functions as obtained by using the standard, Eq.~(\ref{NC-coll}), and 
polynomial, Eq.~(\ref{coll-2}), parameterisations. On the left panel 
we show the results obtained by 
fitting the SIDIS data together with the $A_{12}$ Belle asymmetries 
(both with standard and polynomial parameterisation), while on the right 
panel we show the corresponding results obtained by fitting the SIDIS 
data together with the $A_{0}$ Belle asymmetries.
}
\end{center}
\end{figure}

We have repeated the same fitting procedure as performed with the standard
parameterisation. When fitting the combined SIDIS, $A_{12}^{UL}$ and 
$A_{12}^{UC}$ Belle data, the resulting best fits (not shown) hardly 
exhibit any difference with respect to those obtained with the standard 
parameterisation (Fig.~\ref{fig:Belle-standard-A12fit}). This can be 
seen also from the $\chi^2$'s in Table~\ref{chisq}, where the third line 
is very similar to the first one. As a further confirmation, the 
corresponding best fit plots for ${\cal N}_{\rm fav, dis}^C(z)$, in 
case of the standard and polynomial parameterisations, plotted in 
Fig.~\ref{fig:poly} (left panel) practically coincide up to values of $z$ 
very close to 1.   
 
\begin{figure}[t]
\begin{center}
\includegraphics[width=0.4\textwidth, angle=-90]{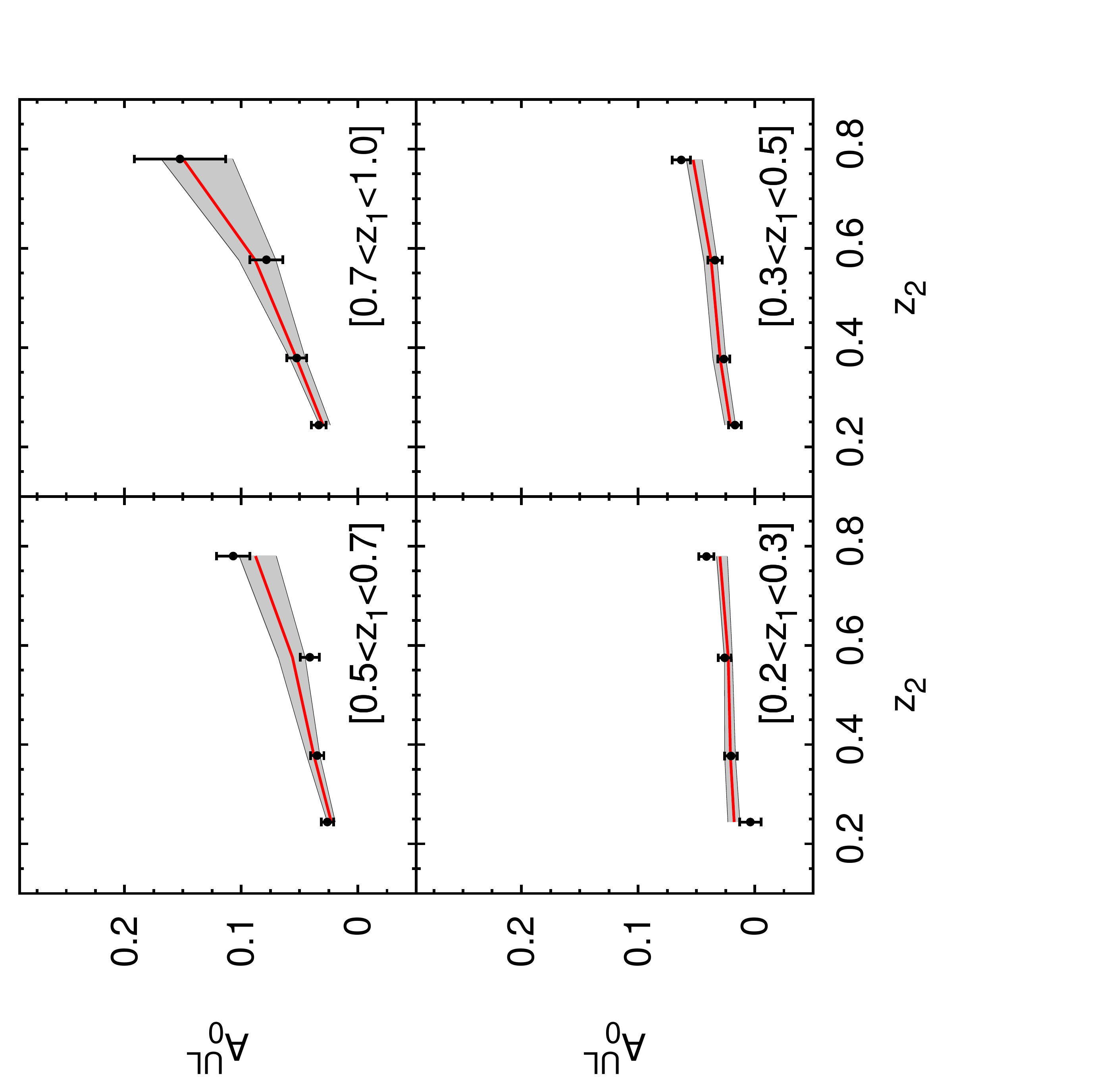}
\hspace*{.3cm}
\includegraphics[width=0.4\textwidth, angle=-90]{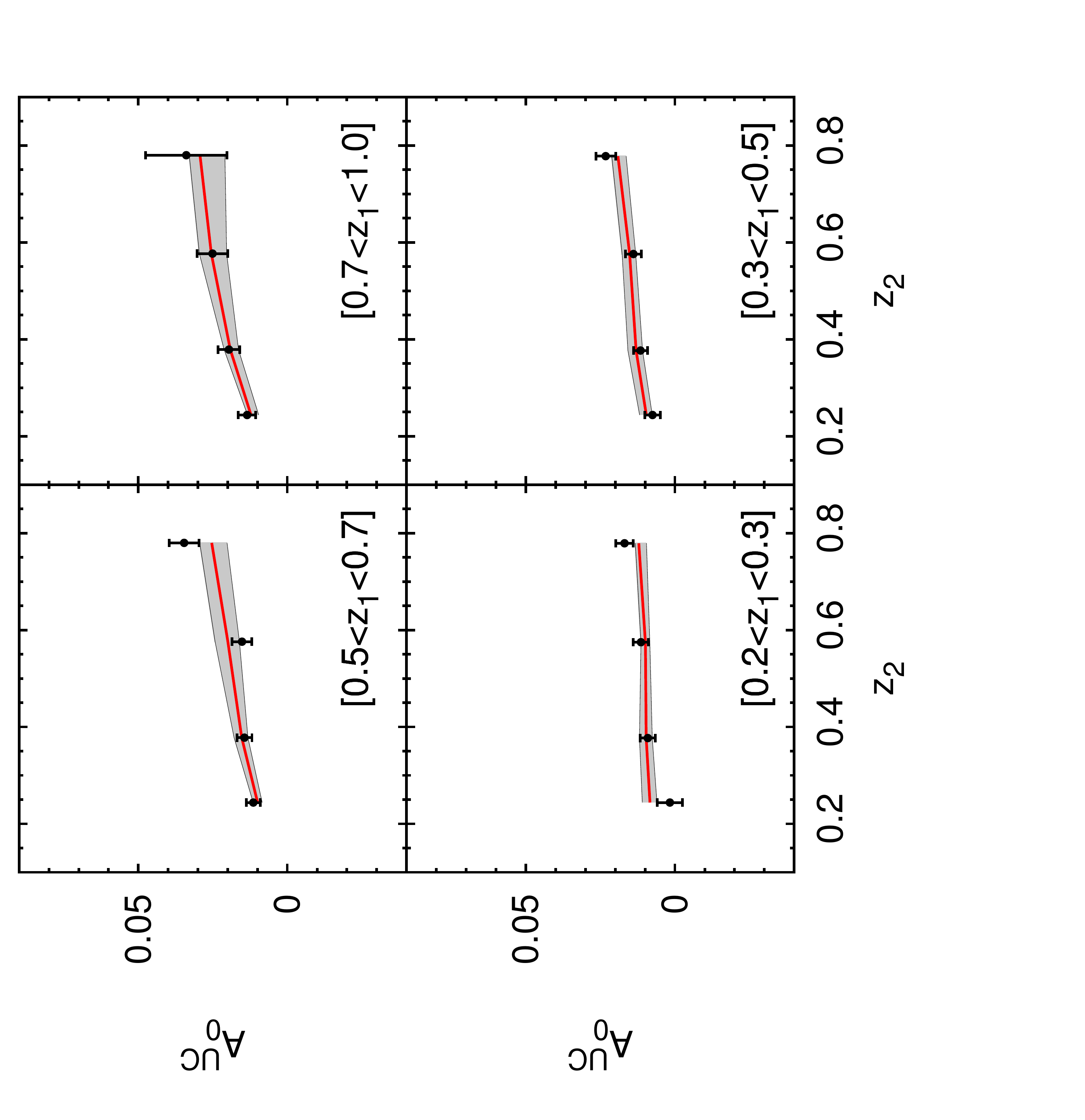}
\\
\includegraphics[width=0.4\textwidth, angle=-90]{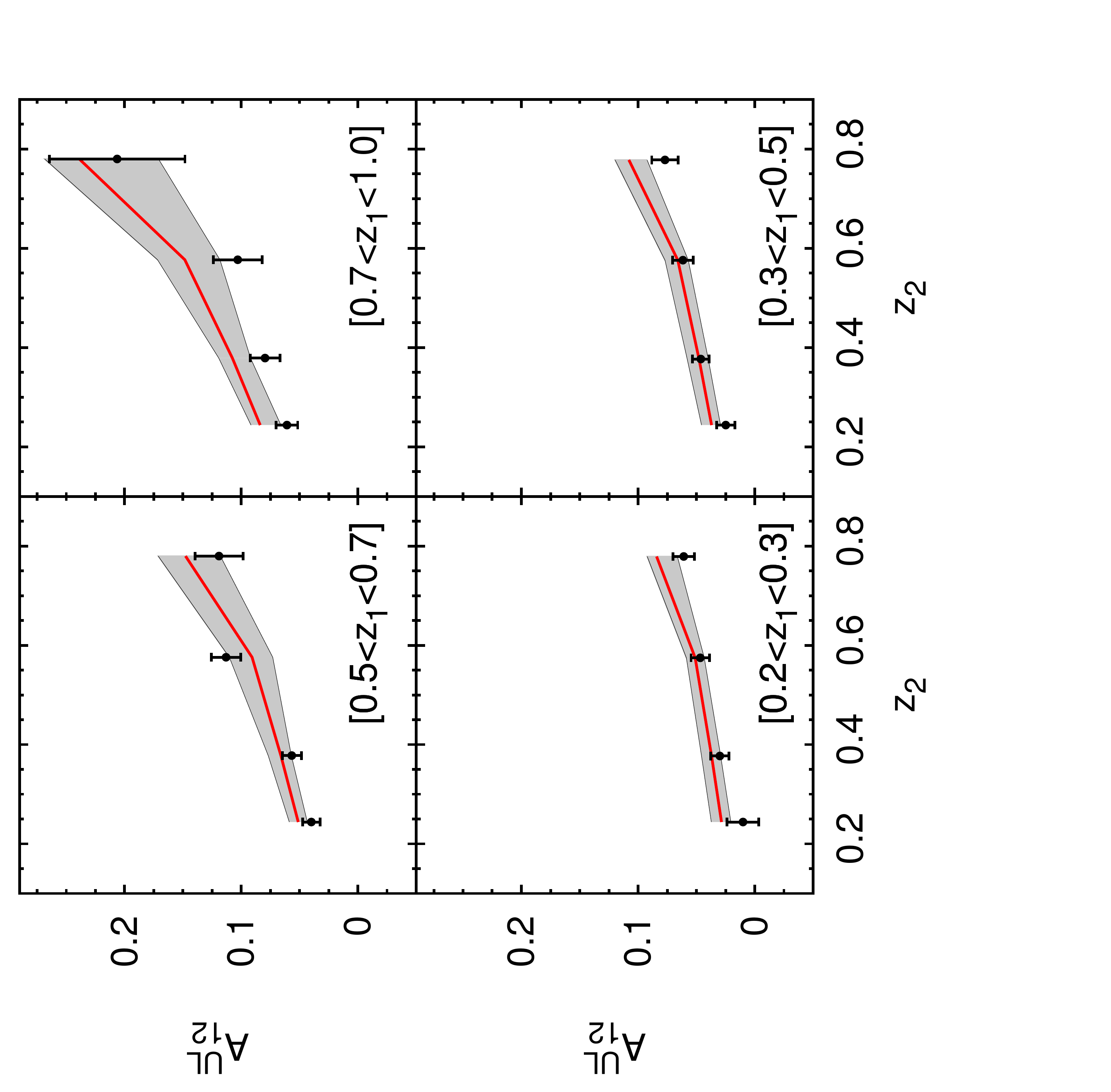}
\hspace*{.3cm}
\includegraphics[width=0.4\textwidth, angle=-90]{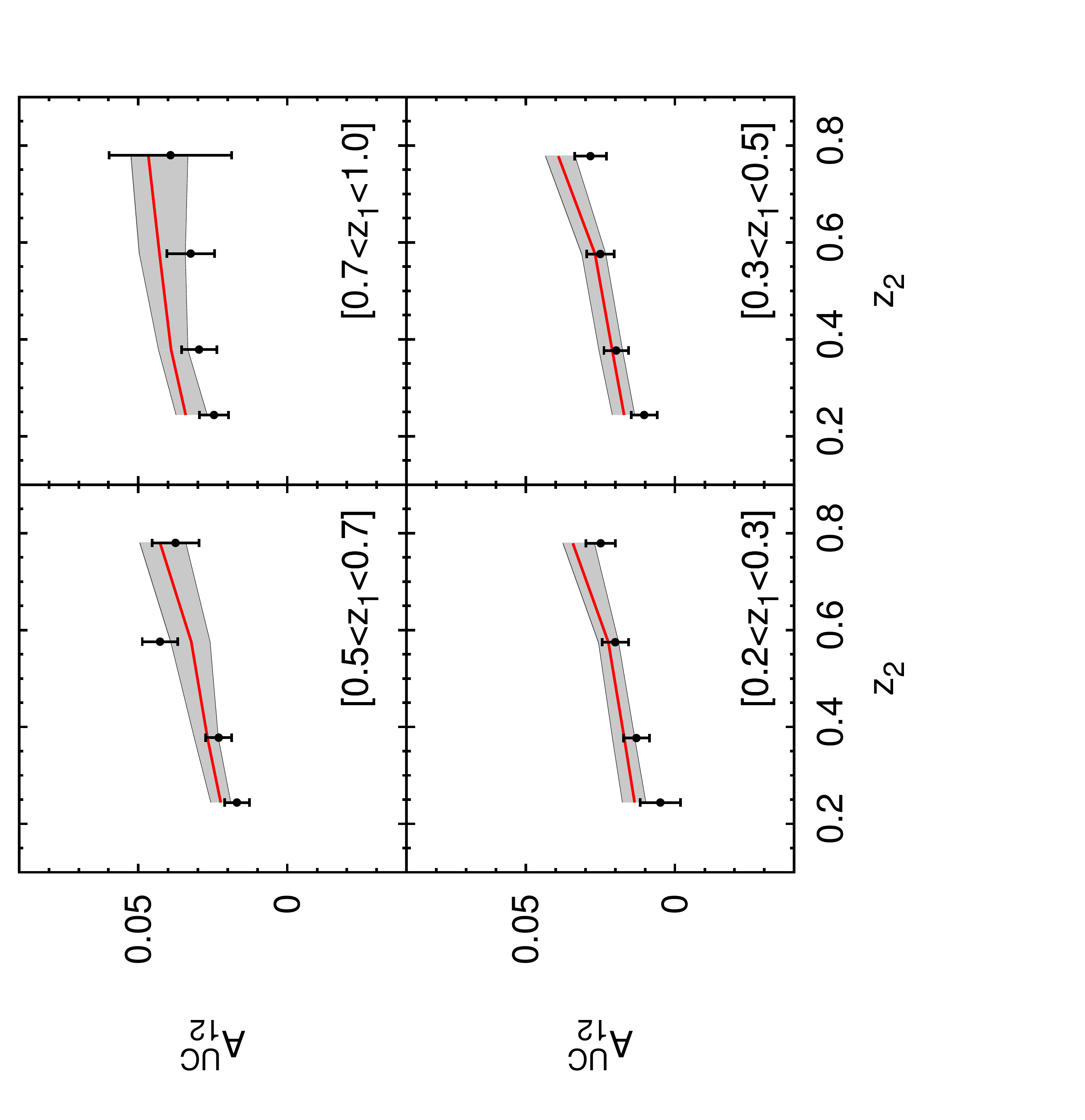}
\vskip -18pt
\caption{\label{fig:Belle-poly-A0fit}
The experimental data on $A_{0}^{UL}$, $A_{0}^{UC}$ (upper plots) and 
$A_{12}^{UL}$ and $A_{12}^{UC}$ (lower plots), as measured by the 
Belle Collaboration~\cite{Seidl:2012er} in unpolarized 
$e^+e^- \to h_1 \, h_2 \, X$ processes, are compared to the curves 
obtained from our global fit. The solid lines correspond to the 
parameters given in Table~\ref{fitpar-poly}, obtained by fitting the 
SIDIS and the $A_{0}$ asymmetries with polynomial parameterisation;
the shaded areas correspond to the statistical uncertainty on the 
parameters, as explained in the text and in Ref.~\cite{Anselmino:2008sga}.
Notice that the $A_{12}^{UL}$ and $A_{12}^{UC}$ data are not included in 
the fit and our curves, with the corresponding uncertainties, are 
simply computed using the parameters of Table~\ref{fitpar-poly}.
}
\end{center}
\end{figure}

The situation is different when best fitting the SIDIS data together 
with $A_{0}^{UL}$ and $A_{0}^{UC}$; in such a case the polynomial 
parameterisation allows a much better best fit, as shown in 
Fig.~\ref{fig:Belle-poly-A0fit}, upper plots. A reasonable agreement 
can also be achieved between the data and the computed values of 
$A_{12}^{UL}$ and $A_{12}^{UC}$, as shown by the $\chi ^2$ values 
in Table~\ref{chisq} and by the lower plots in Fig.~\ref{fig:Belle-poly-A0fit}. 
In this case the polynomial form of ${\cal N}^{\C}_{\rm fav, dis}(z)$
differs from the standard one, as shown in the right plots in 
Fig.~\ref{fig:poly}.

Notice, again, that the large $\chi ^2$ values of the computed $A_{12}^{UL}$ 
is almost completely due to the last $z$ bins, which correspond to the quasi exclusive region. Also, the larger $\chi ^2$ values corresponding to SIDIS 
data are mainly due to a slightly worse description of HERMES $\pi^-$ azimuthal moments. The values of the parameters obtained using the polynomial shape 
of ${\cal N}^{\C}_{\rm fav, dis}(z)$, Eq.~(\ref{coll-2}), are given in 
Table~\ref{fitpar-poly}.

\begin{table}[t]
\caption{
Best values of the 9 free parameters fixing the $u$ and $d$ quark
transversity distribution functions and the favoured and
disfavoured Collins fragmentation functions, as obtained by fitting 
simultaneously SIDIS data on the Collins asymmetry and Belle data on 
$A_{0}^{UL}$ and $A_{0}^{UC}$. The transversity distributions 
are parameterised according to Eqs.~(\ref{tr-funct}), (\ref{NT-transv}) 
and the Collins fragmentation functions according to the polynomial 
parameterisation, Eqs.~(\ref{coll-funct}), (\ref{coll-2}) and 
(\ref{hpcollins}). We obtain a total $\chi^2/{\rm d.o.f.} = 1.01$.
The statistical errors quoted for each parameter correspond to the shaded 
uncertainty areas in Fig.~\ref{fig:Belle-poly-A0fit}, as explained in the 
text and in the Appendix of Ref.~\cite{Anselmino:2008sga}. 
\label{fitpar-poly}}
\vskip 18pt
\renewcommand{\tabcolsep}{0.4pc} 
\renewcommand{\arraystretch}{1.2} 
\begin{tabular}{@{}ll}
 \hline
 $N_{u}^{\T}$ = $0.36^{+0.19}_{-0.12}$ & $N_{d}^{\T}$ = $ -1.00^{+0.40}_{0.00}$      \\
 $\alpha$ =  $1.06^{+0.87}_{-0.56}$ & $\beta$  = $3.66^{+5.87}_{-2.78}$ \\
 \hline
 $N_{\rm fav}^{\C}$  = $1.00^{+0.00}_{-0.36}$ & $N_{\rm dis}^{\C}$ = 
 $-1.00^{+0.19}_{-0.00}$ \\
 $a$  = $-2.36^{+1.24}_{-0.98}$  & $b$   = $2.12^{+0.61}_{-1.12}$    \\
 $M^2_h = 0.67^{+1.09}_{-0.36}$ GeV$^2$ &\\
 \hline
\end{tabular}
\end{table}

\subsection{The extracted transversity and Collins functions; predictions and final comments}

\begin{figure}[t]
\begin{center}
\includegraphics[width=0.5\textwidth, angle=-90]
{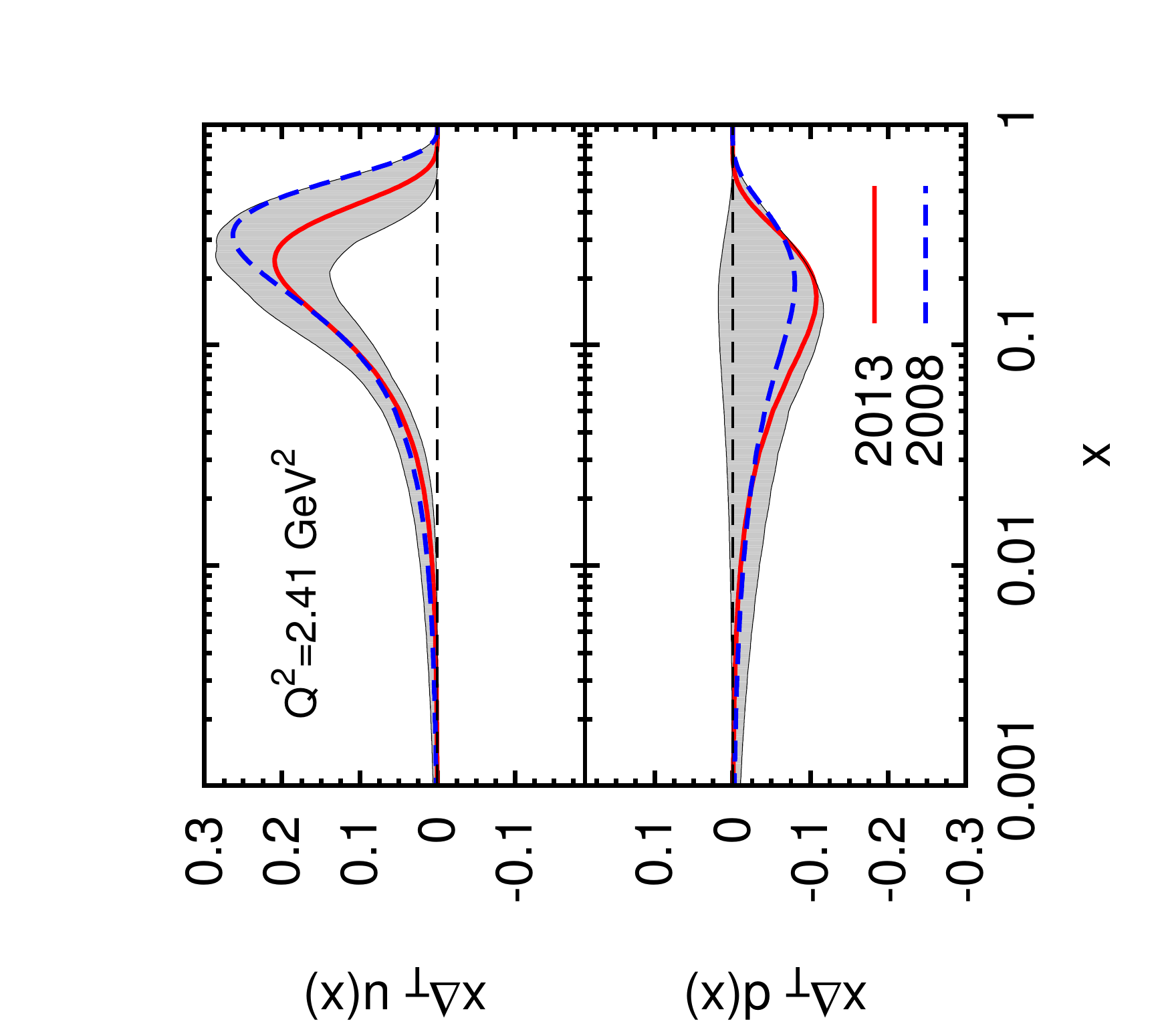}
\hspace*{.3cm}
\includegraphics[width=0.5\textwidth,angle=-90]
{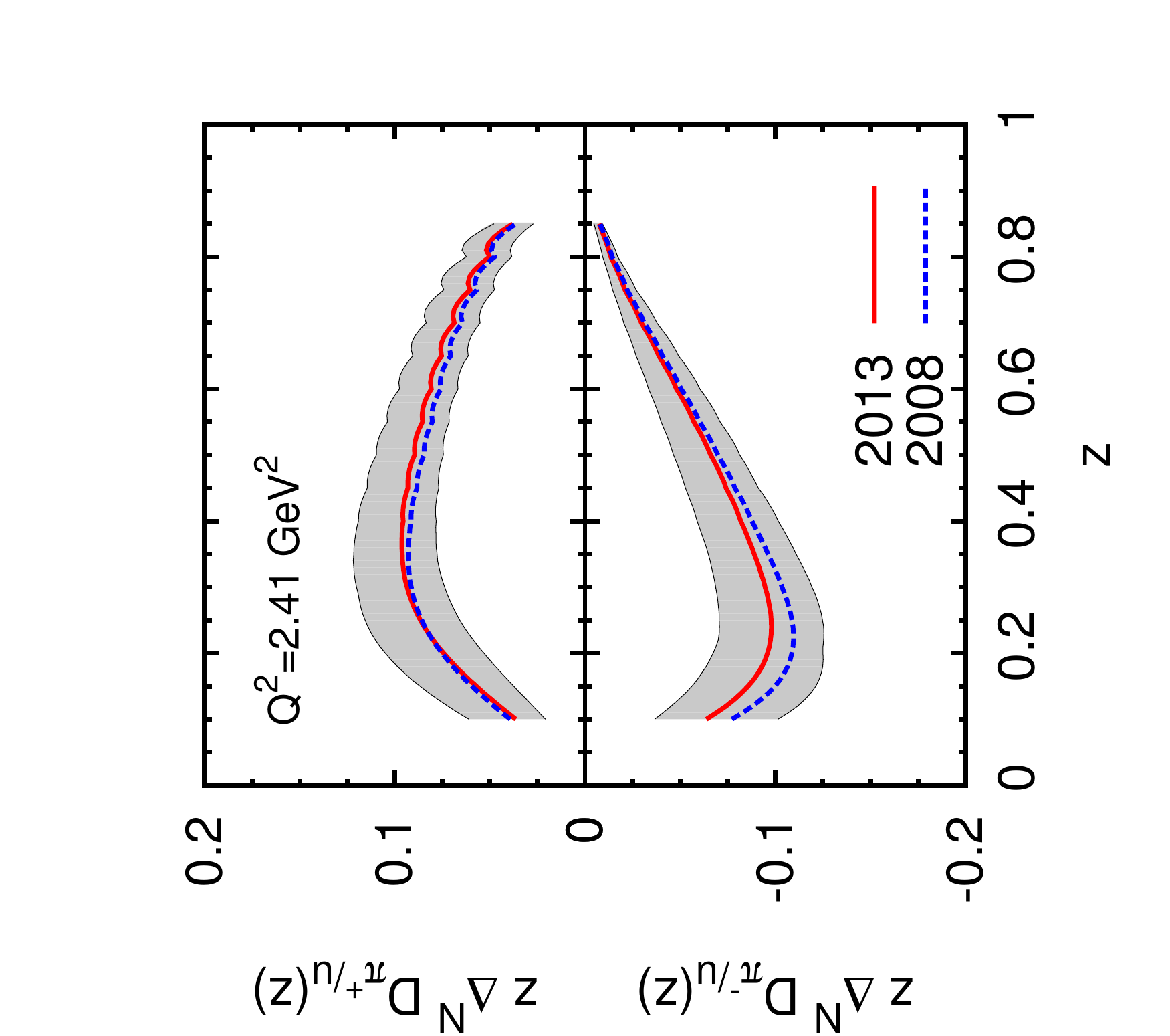}
\caption{\label{fig:newh1-collins-A12}
In the left panel we plot (solid red lines) the transversity 
distribution functions $x \, h_{1q}(x) = x \, \Delta_T q(x)$ for $q = u,d$,
with their uncertainty bands (shaded areas), obtained from our best fit 
of SIDIS data on $A_{UT}^{\sin(\phi_h + \phi_S)}$ and $e^+e^-$ data on 
$A_{12}$, adopting the standard parameterisation (Table~\ref{fitpar}). 
Similarly, in the right panel we plot the corresponding first moment of 
the favoured and disfavoured Collins functions, Eq.~(\ref{coll-mom2}). All 
results are given at $Q^2=2.41$ GeV$^2$. 
The dashed blue lines show the same quantities as obtained in 
Ref.~\cite{Anselmino:2008jk} using the data then available on 
$A_{UT}^{\sin(\phi_h + \phi_S)}$ and $A_{12}^{UL}$. 
}
\end{center}
\end{figure}

\begin{figure}[t]
\begin{center}
\includegraphics[width=0.5\textwidth, 
angle=-90]{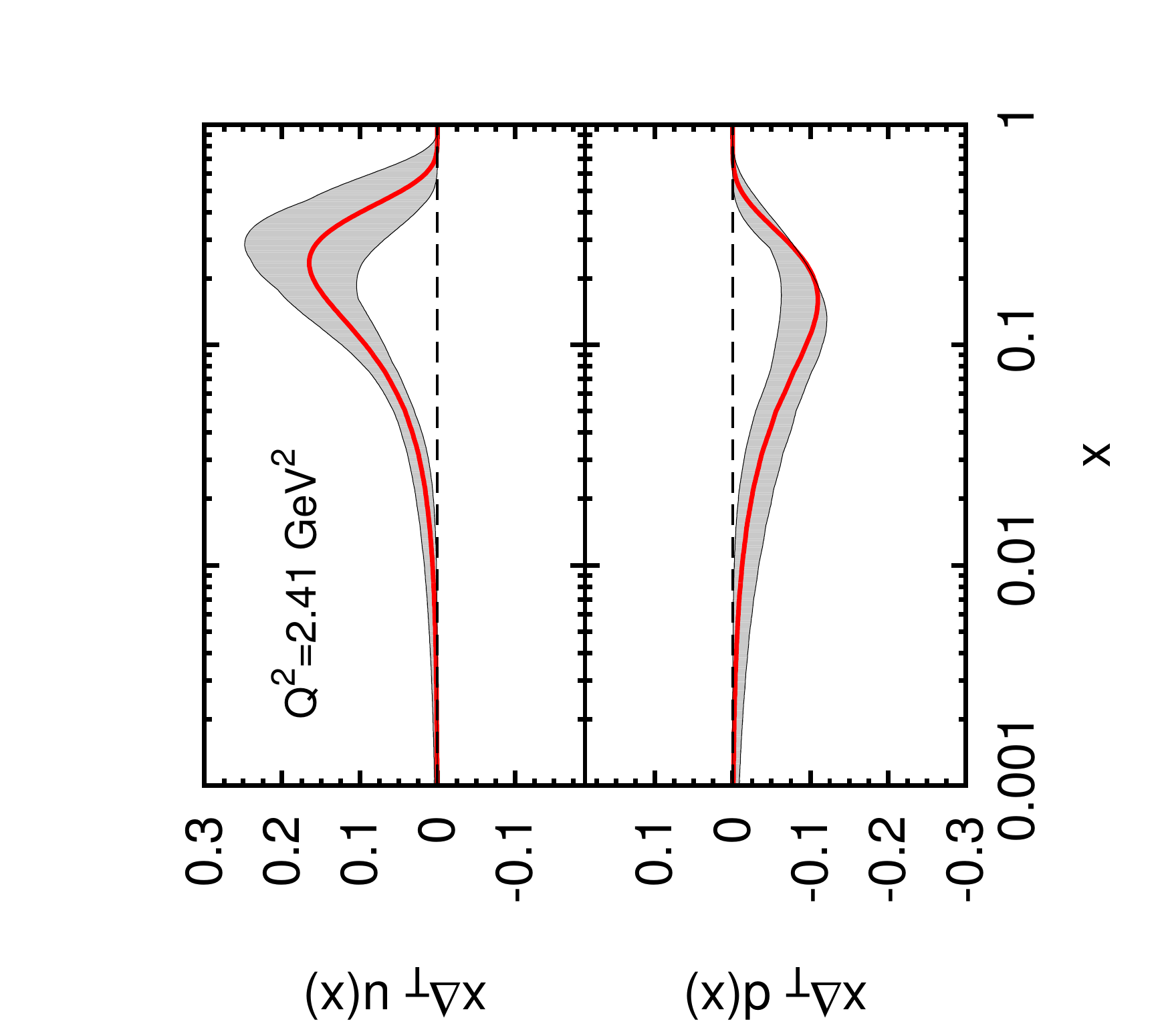}
\hspace*{.3cm}
\includegraphics[width=0.5\textwidth, 
angle=-90]{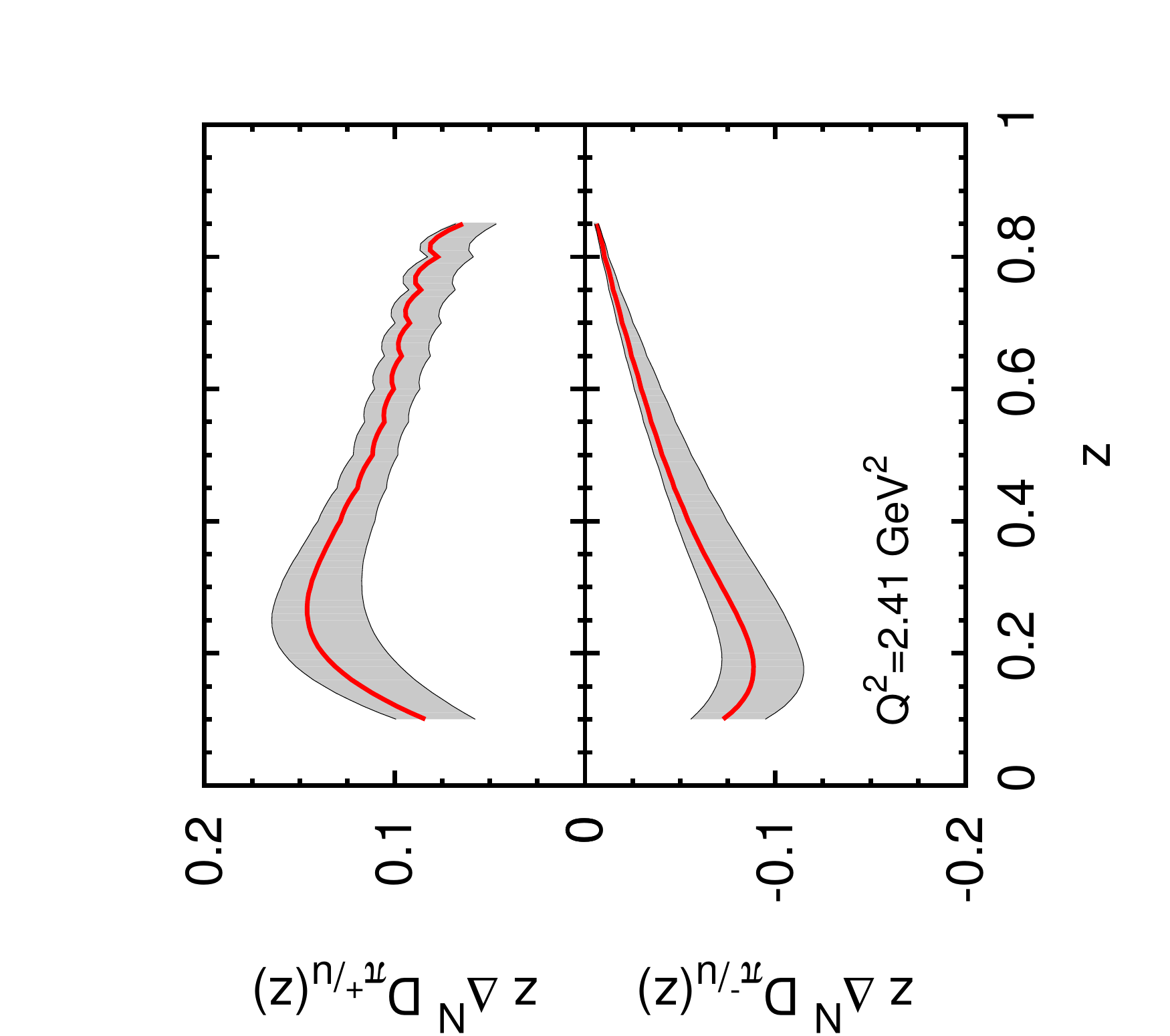}
\caption{\label{fig:newh1-collins-A0}
In the left panel we plot (solid red lines) the transversity 
distribution functions $x \, h_{1q}(x) = x \, \Delta_T q(x)$ for $q = u,d$,
with their uncertainty bands (shaded areas), obtained from our best fit 
of SIDIS data on $A_{UT}^{\sin(\phi_h + \phi_S)}$ and $e^+e^-$ data on 
$A_{0}$, adopting the polynomial parameterisation (Table~\ref{fitpar-poly}). 
Similarly, in the right panel we plot the corresponding first moment of 
the favoured and disfavoured Collins functions, Eq.~(\ref{coll-mom2}). All 
results are given at $Q^2=2.41$ GeV$^2$. 
}
\end{center}
\end{figure}

Our newly extracted transversity and Collins functions are shown in
Figs.~\ref{fig:newh1-collins-A12} and \ref{fig:newh1-collins-A0}; to be 
precise, in the left panels we show $x \,\Delta_T q(x) = x \, h_{1q}(x)$, 
for $u$ and $d$ quarks, while in the right panels we plot: 
\be
z\,\Delta ^N\! D _{h/q^\ua}(z)= z \!\! \int d^2\bpp \Delta ^N\!
D_{h/q^\ua} (z,\pp) = z \!\! \int d^2\bpp \frac{2\,\pp}{z\,m_h} \> 
H_1^{\perp q}(z, \pp) = 4z \, H_1^{\perp (1/2)q}(z) \label{coll-mom2}
\ee
for $h = \pi^\pm$ and $q = u$. The Collins results for $d$ quark are 
not shown explicitly, but could be obtained from Tables~\ref{fitpar}
and \ref{fitpar-poly}.   

Fig.~\ref{fig:newh1-collins-A12} shows the results which best fit the 
COMPASS and HERMES SIDIS data on $A_{UT}^{\sin(\phi_h + \phi_S)}$, 
together with the Belle results on $A_{12}^{UL}$ and $A_{12}^{UC}$, 
using the standard parameterisation. The red solid lines correspond 
to the parameters given in Table~\ref{fitpar}. The shaded bands show 
the uncertainty region, which is the region spanned by the 1500 
different sets of parameters fixed according to the procedure explained 
above and in the Appendix of Ref.~\cite{Anselmino:2008sga}. The blue 
dashed lines show, for comparison, our previous 
results~\cite{Anselmino:2008jk}: the difference between the solid red 
and dashed blue lines is only due to the updated SIDIS and $A_{12}^{UL}$ 
data used here, with the addition of $A_{12}^{UC}$, while keeping the 
same parameterisation. The present and previous results agree within the 
uncertainty band: one could at most notice a slight decrease of the 
new $u$ quark transversity distribution at large $x$ values.    

Fig.~\ref{fig:newh1-collins-A0} shows the results which best fit the 
COMPASS and HERMES SIDIS data on $A_{UT}^{\sin(\phi_h + \phi_S)}$, 
together with the Belle results on $A_{0}^{UL}$ and $A_{0}^{UC}$, 
using the polynomial parameterisation. The red solid lines correspond 
to the parameters given in Table~\ref{fitpar-poly}. This is not a simple 
updating of our previous 2008 fit~\cite{Anselmino:2008jk}, as we use 
different sets of data (SIDIS and $A_{0}$ rather than SIDIS 
and $A_{12}$) with a different polynomial parameterisation. In this case 
the comparison with the 2008 results is less significant. If comparing the 
results of Fig.~\ref{fig:newh1-collins-A12} and \ref{fig:newh1-collins-A0},    
one notices a sizeable difference in the favoured $(u/\pi^+$) Collins 
function, and less evident differences in the transversity distributions.  

\begin{figure}[t]
\begin{center}
\includegraphics[width=0.9\textwidth, angle=0]{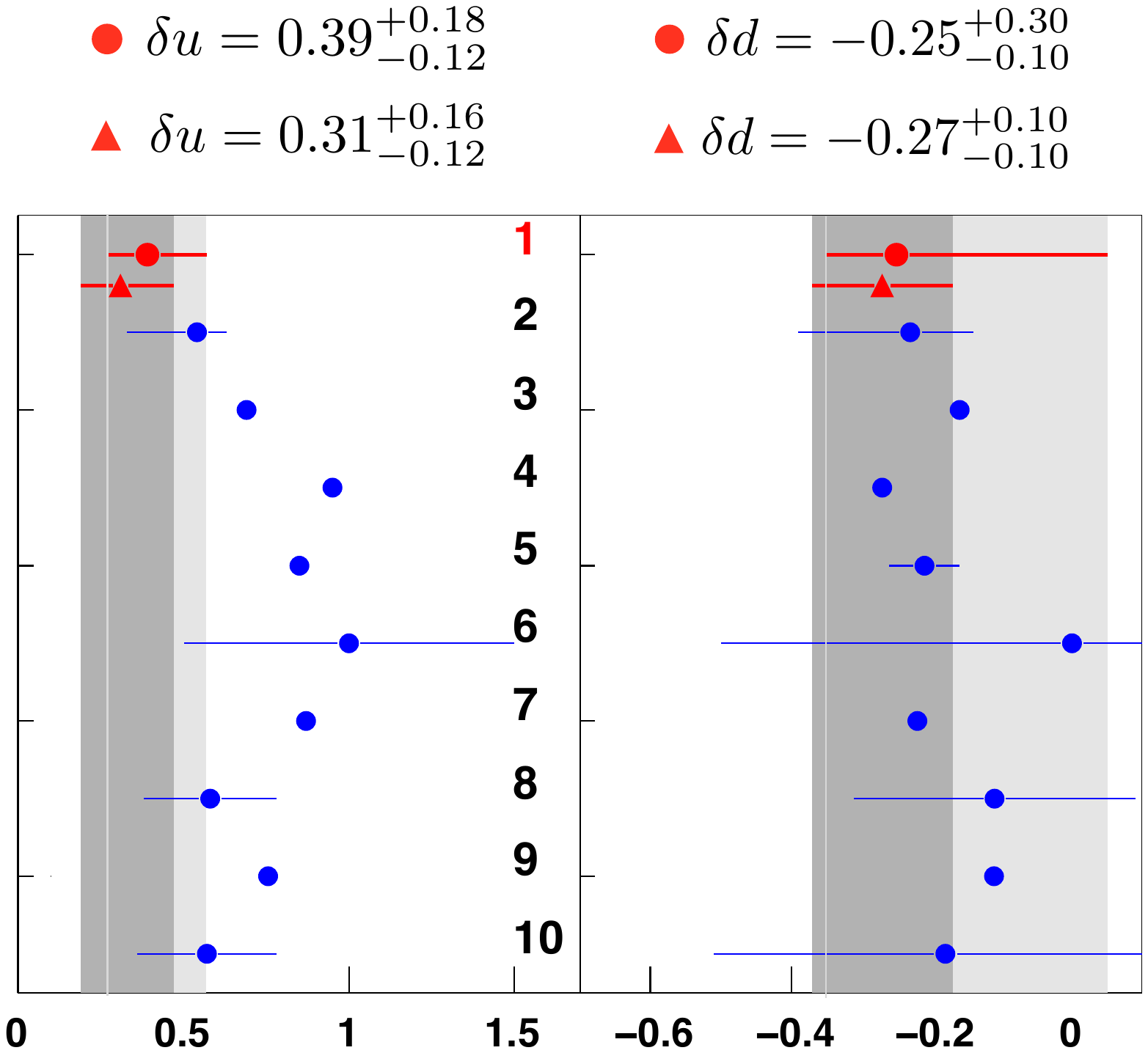}
\vskip -36pt
\caption{\label{fig:tensorcharge}
The tensor charge $\delta q \equiv \int_0^1 dx \, [\Delta_T q(x) - 
\Delta_T {\bar q}(x)]$
for $u$ (left) and $d$ (right) quarks, computed using the transversity 
distributions obtained from our best fits, Table~\ref{fitpar} (top solid red 
circles) and Table~\ref{fitpar-poly} (solid red triangles). The gray areas 
correspond to the statistical uncertainty bands in our extraction. These 
results are compared with those given in Ref.~\cite{Anselmino:2008jk} 
(number 2) and with the results of the model calculations of 
Refs.~\cite{Cloet:2007em,Wakamatsu:2007nc,Gockeler:2005cj,He:1994gz,
Pasquini:2006iv,Gamberg:2001qc,Hecht:2001ry,Bacchetta:2012ty}
(respectively, numbers 10 and 3-9).}
\end{center}
\end{figure}

In Fig.~\ref{fig:tensorcharge} we show, for comparison with similar results presented in Ref.~\cite{Anselmino:2008jk}, the tensor charge, corresponding 
to our best fit transversity distributions, as given in Tables~\ref{fitpar}
and \ref{fitpar-poly}. Our extracted values are shown at $Q^2 = 0.8$ GeV$^2$ 
and compared with several model computations. One should keep in mind that 
our estimates are based on the assumption of a negligible contribution from 
sea quarks and on a set of data which still cover a limited range of $x$ 
values.    

All other results are shown at the scale $Q^2 = 2.41$ GeV$^2$. The evolution
to the chosen value has been obtained by evolving at LO the collinear part
of the factorized distribution and fragmentation functions. The TMD 
evolution, which might affect the $\kt$ and $\pp$ dependence, is not
yet known for the Collins function. Consistently, it has not been taken 
into account for the other distribution and fragmentation functions.      
 
\begin{figure}[t]
\begin{center}
\includegraphics[width=0.26\textwidth, angle=-90]{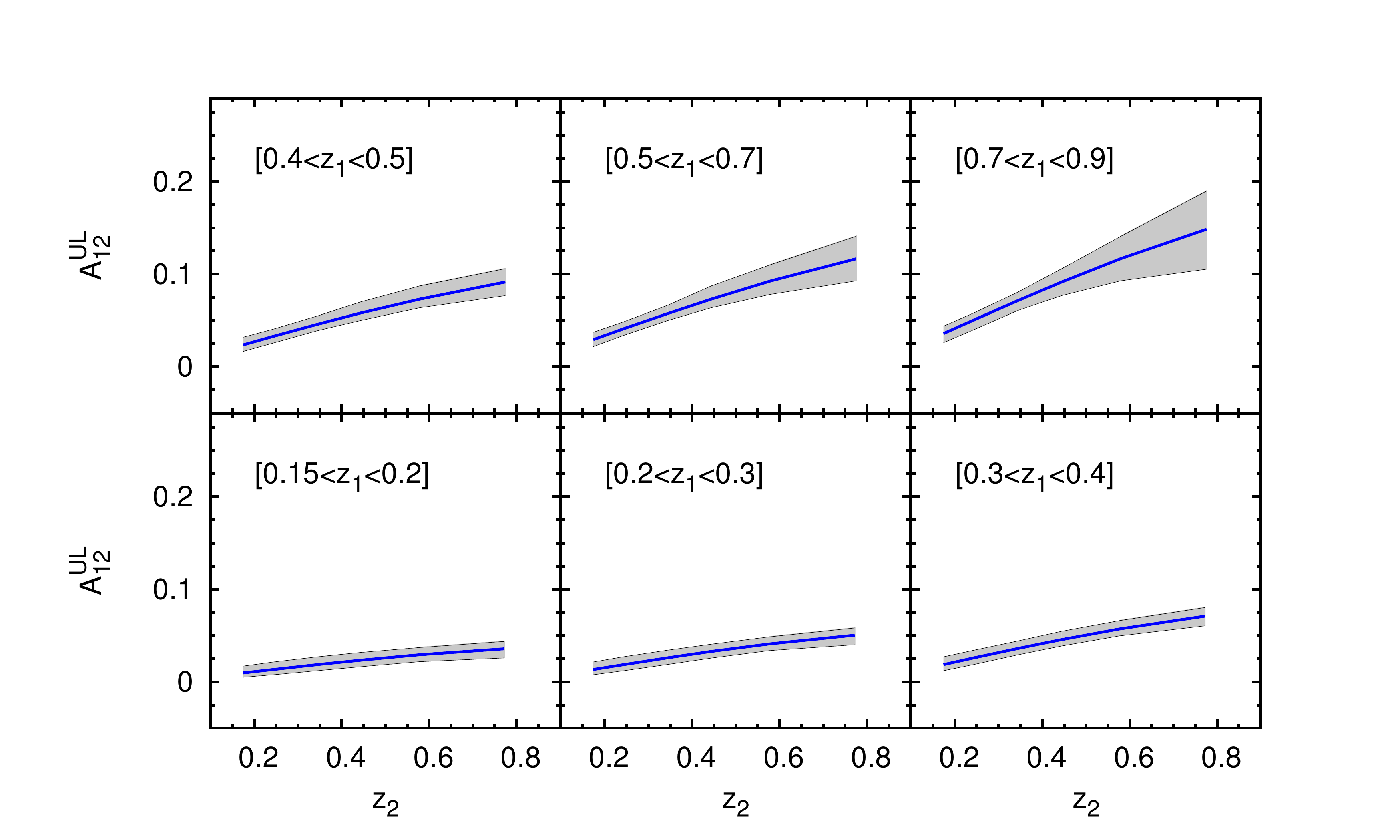}
\hspace*{.2cm}
\includegraphics[width=0.26\textwidth, angle=-90]{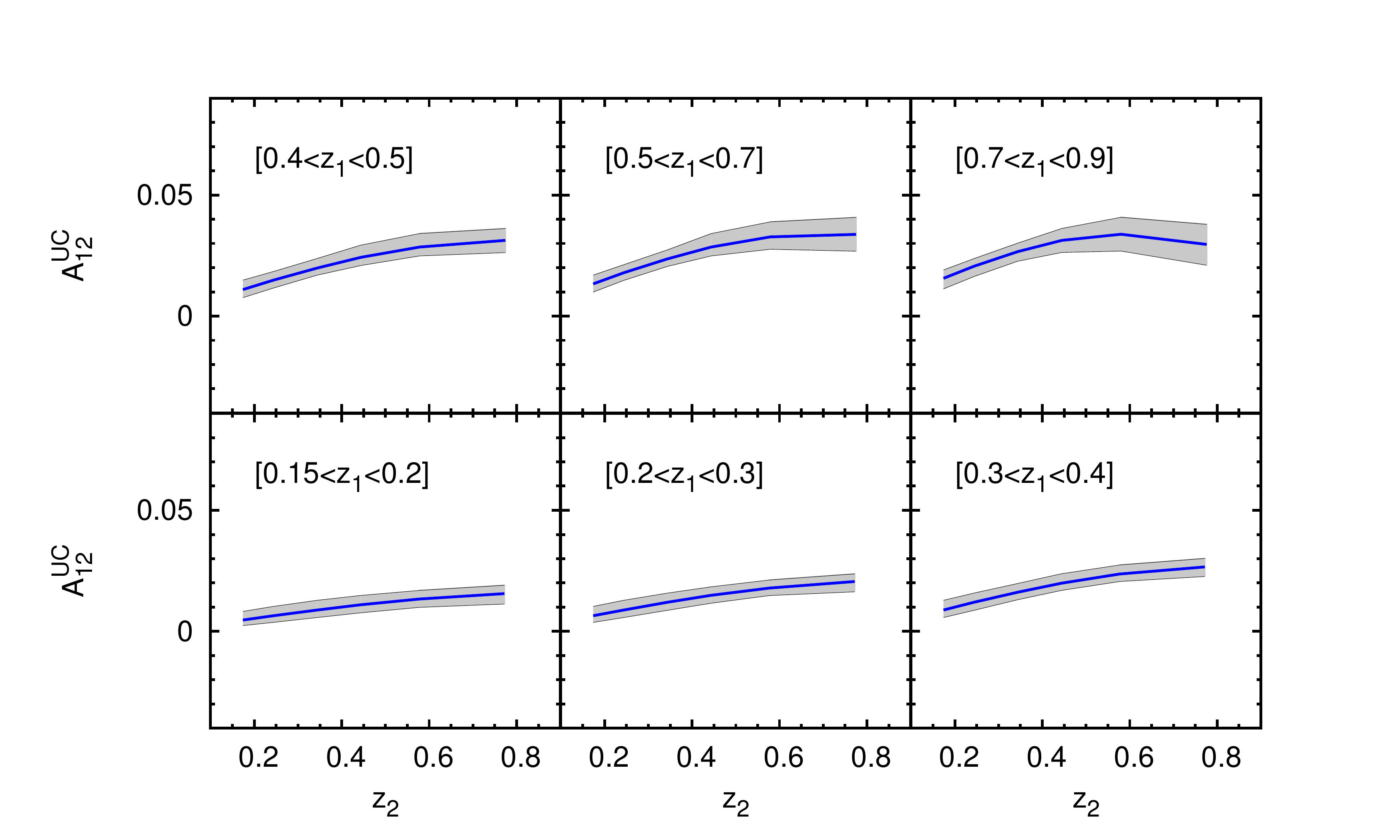}
\\
\includegraphics[width=0.26\textwidth, angle=-90]{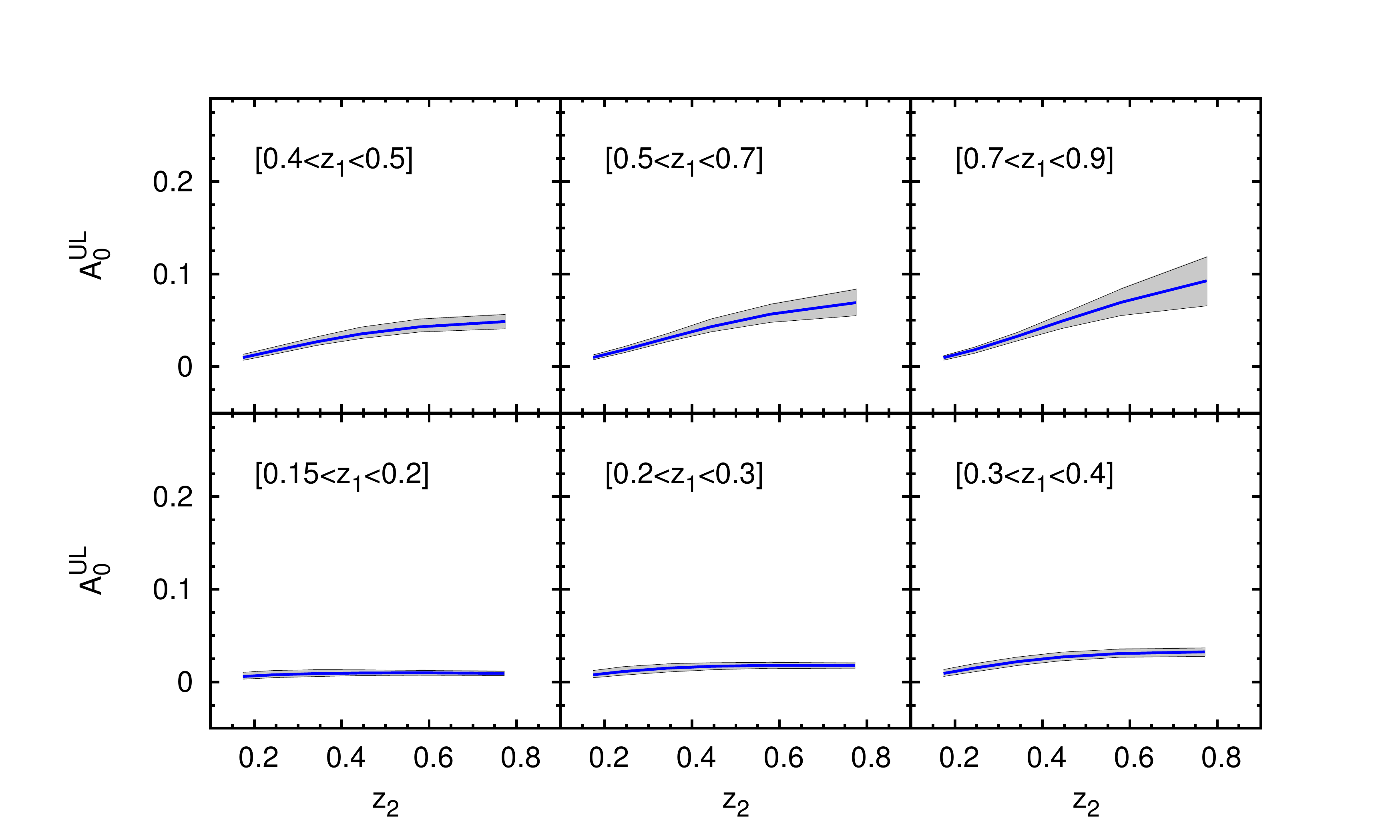}
\hspace*{.2cm}
\includegraphics[width=0.26\textwidth, angle=-90]{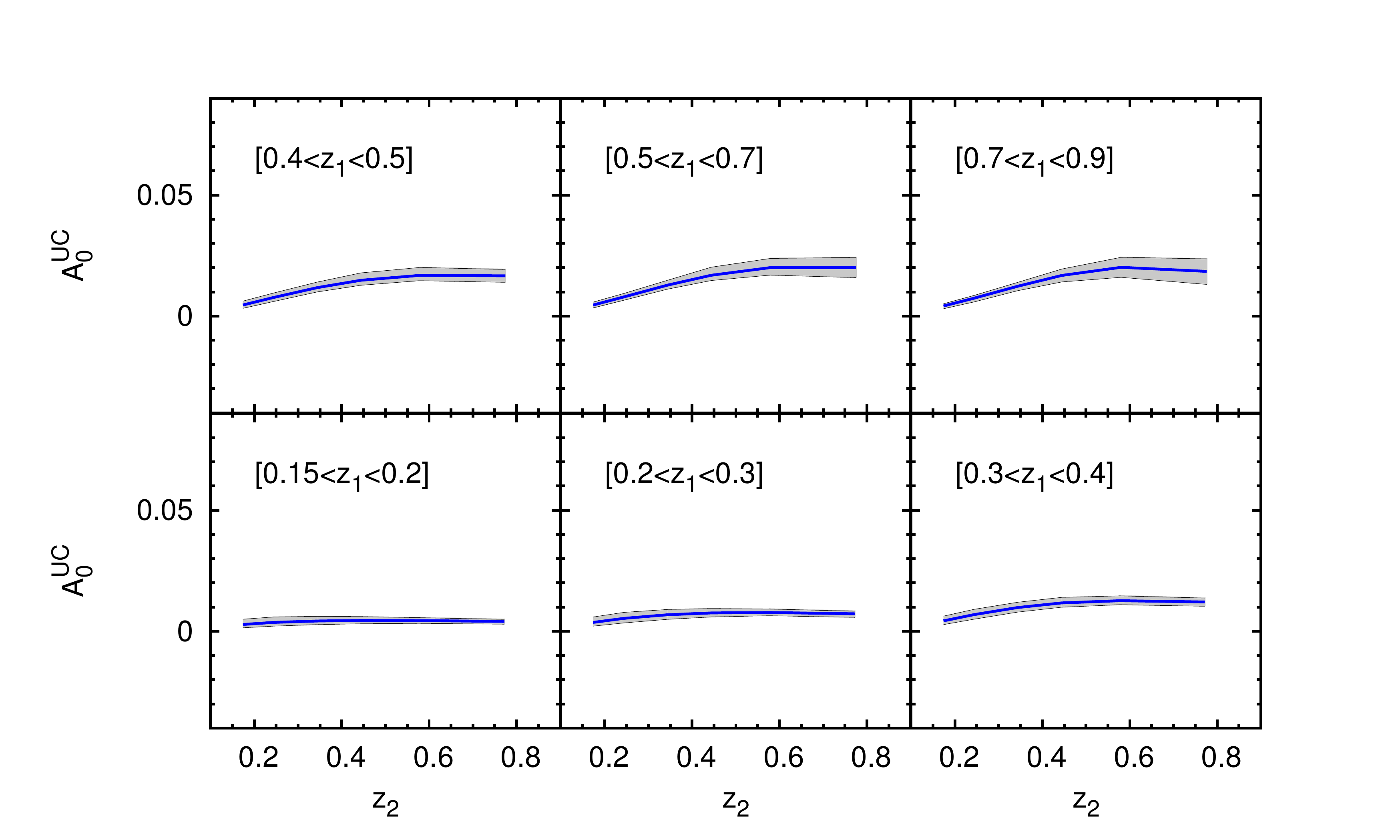}
\caption{\label{fig:babar-A12-stand}
Estimates, obtained from our global fit, for the azimuthal correlations 
$A_{12}^{UL}$, $A_{12}^{UC}$, $A_{0}^{UL}$ and $A_{0}^{UC}$ in unpolarized 
$e^+e^- \to h_1 h_2 \, X$ processes at BaBar~\cite{Garzia:2012za}. 
The solid lines correspond to the parameters given in Table~\ref{fitpar}, 
obtained by fitting the $A_{12}$ Belle asymmetry; the shaded area 
corresponds to the uncertainty on these parameters, as explained in the text.
}
\end{center}
\end{figure}

\begin{figure}[t]
\begin{center}
\includegraphics[width=0.26\textwidth, angle=-90]{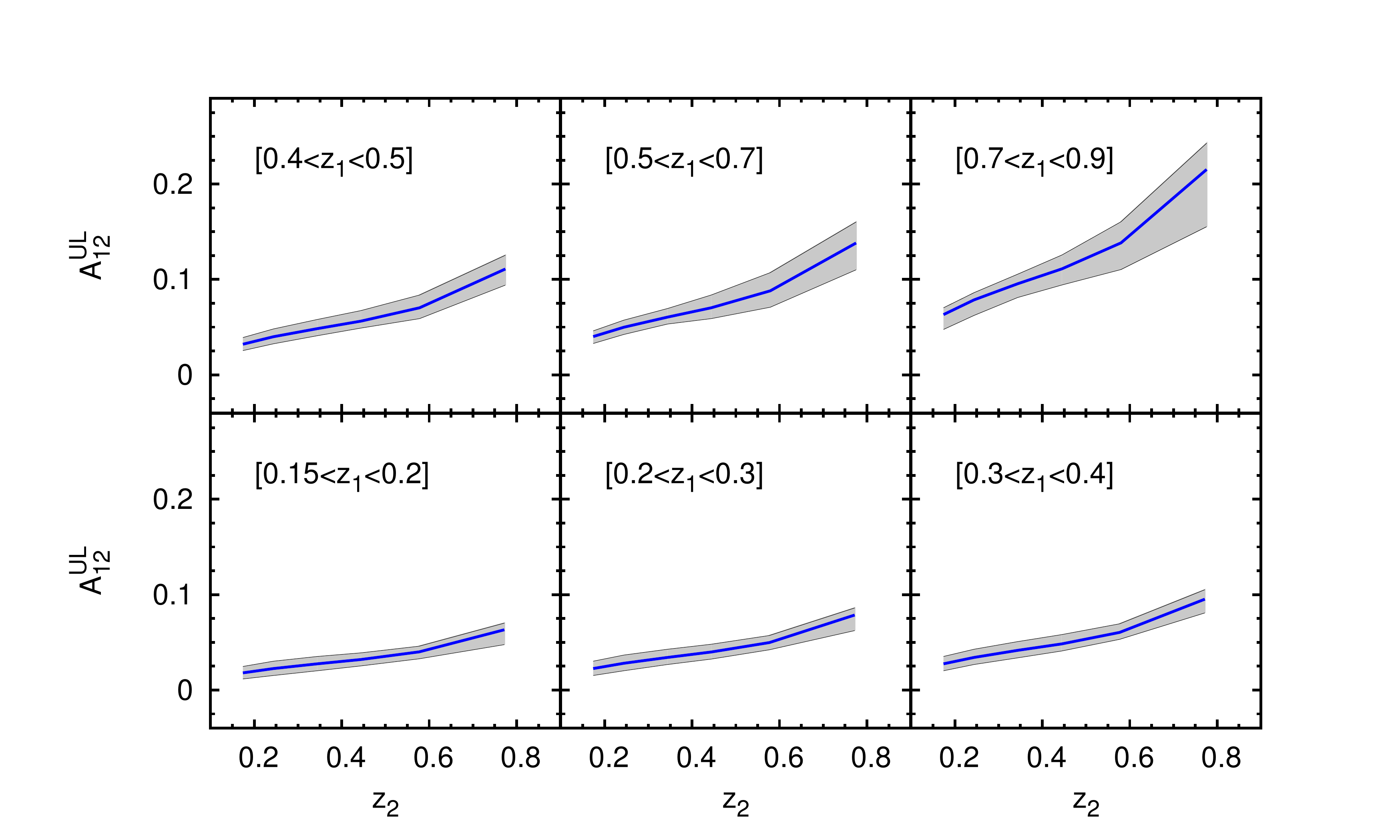}
\hspace*{.2cm}
\includegraphics[width=0.26\textwidth, angle=-90]{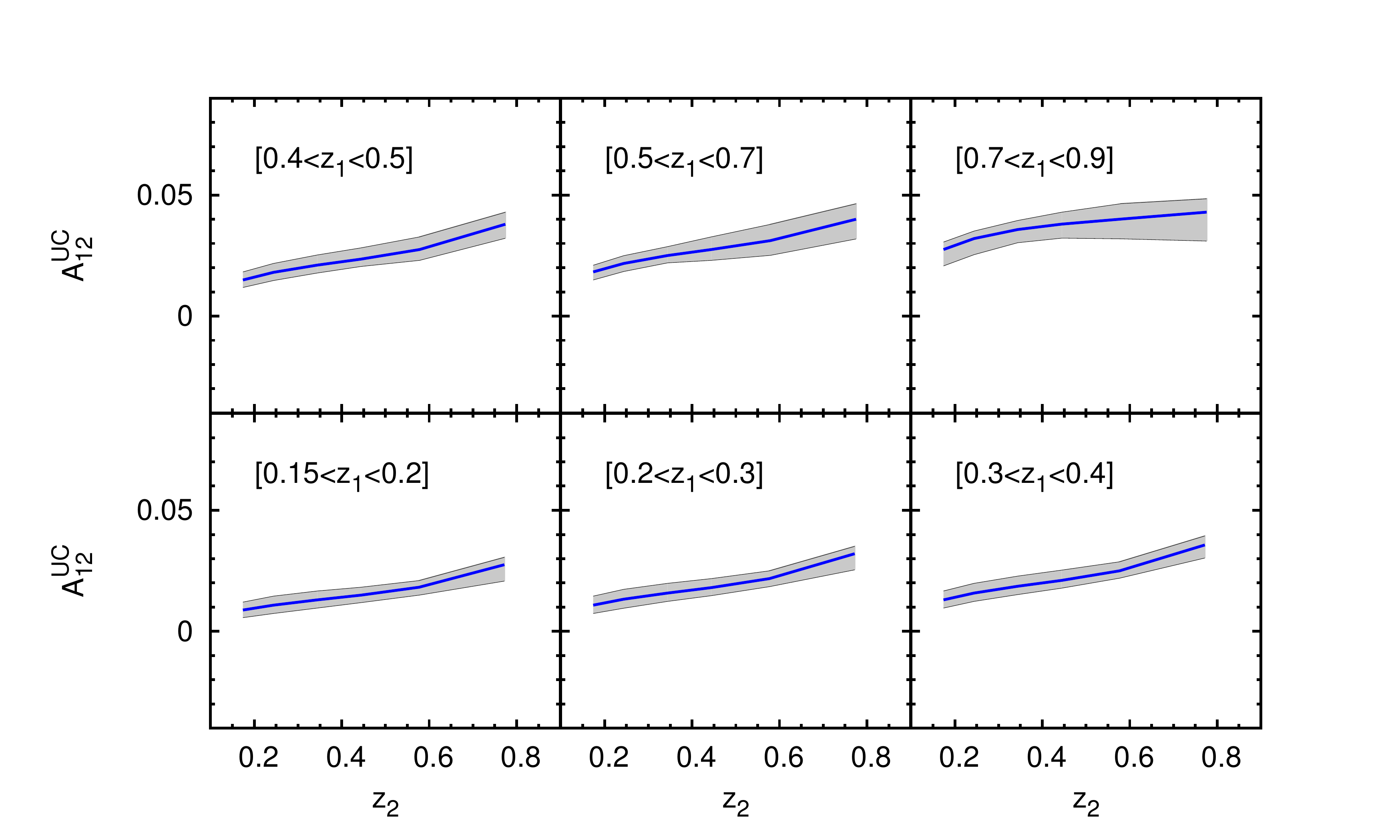}
\\
\includegraphics[width=0.26\textwidth, angle=-90]{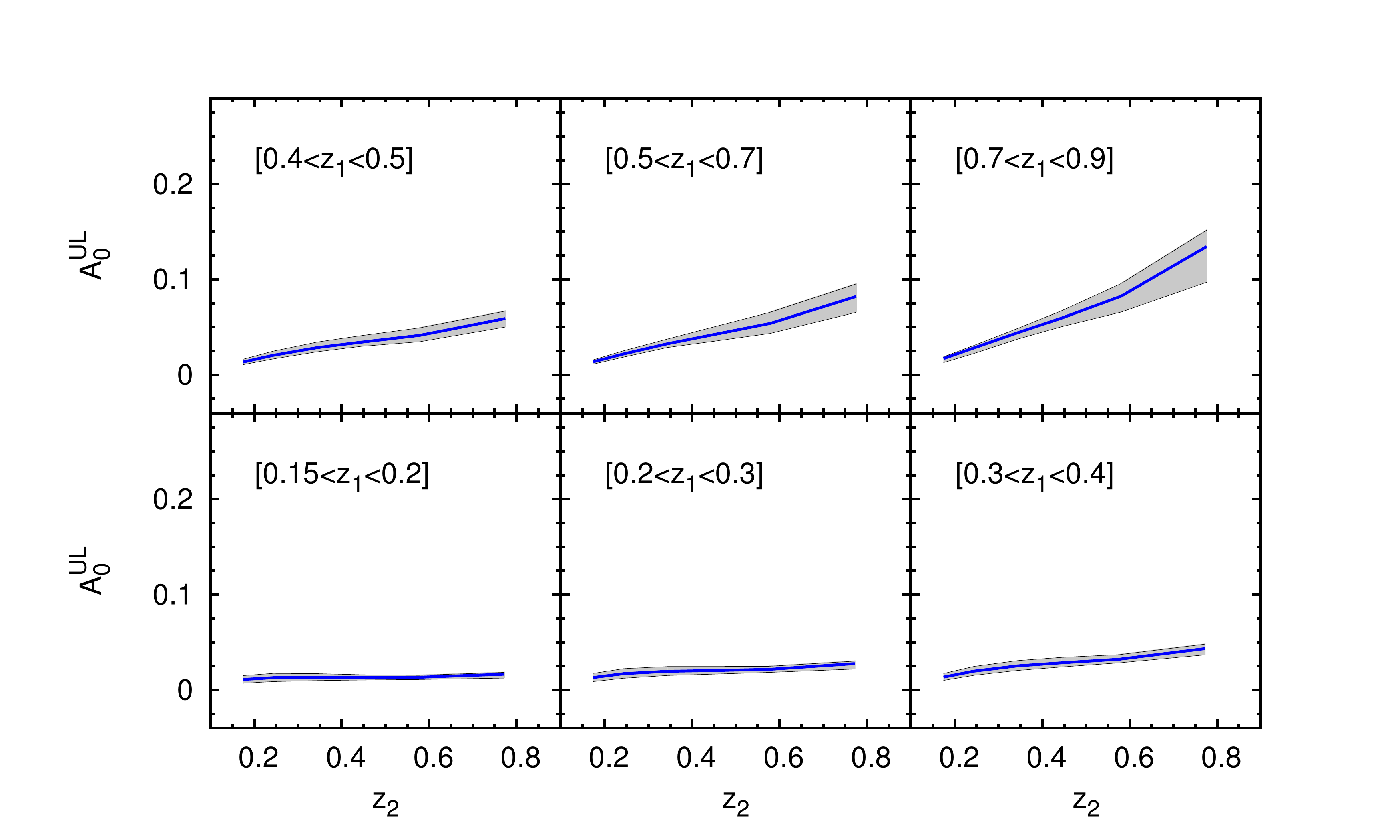}
\hspace*{.2cm}
\includegraphics[width=0.26\textwidth, angle=-90]{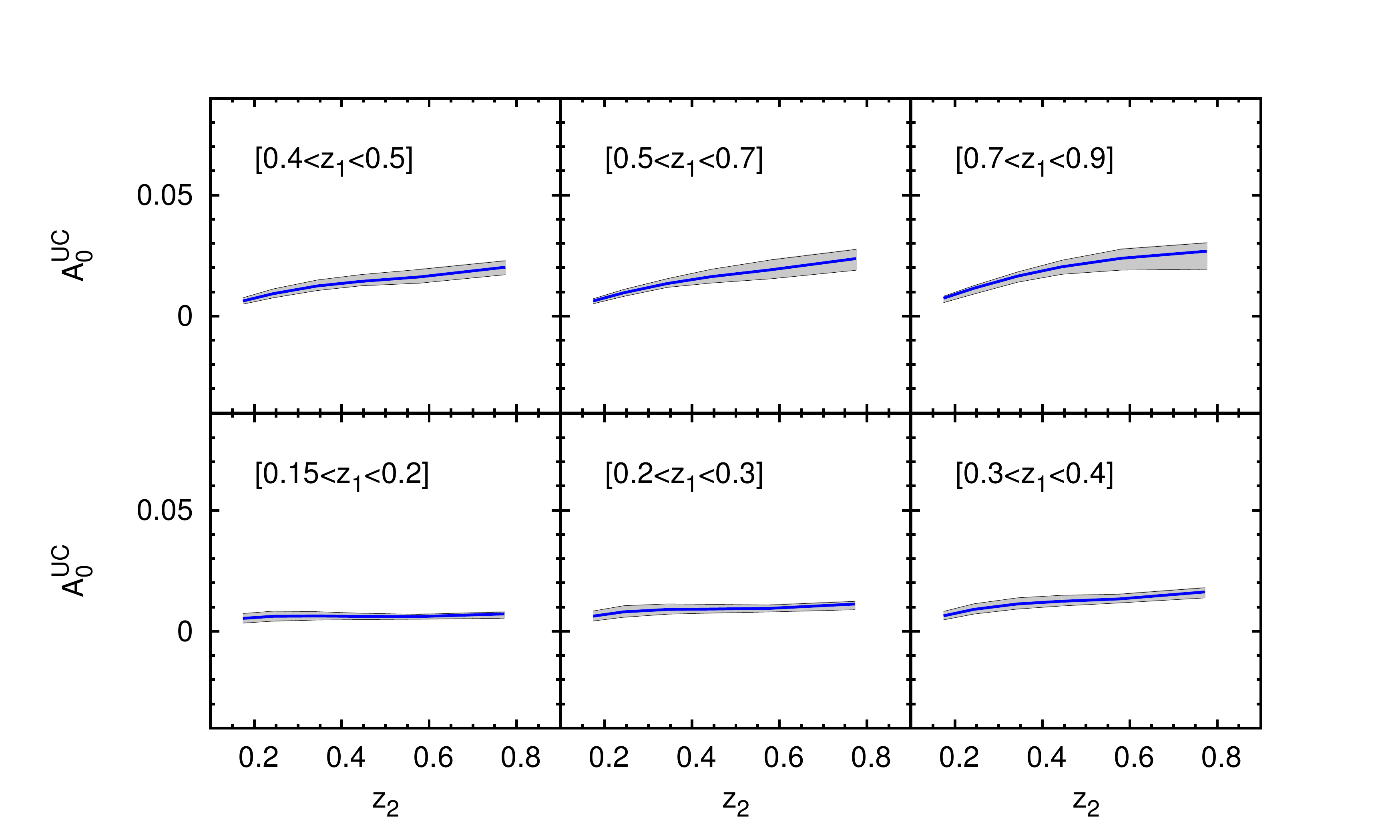}
\caption{\label{fig:babar-A0-poly}
Estimates, obtained from our global fit, for the azimuthal correlations 
$A_{12}^{UL}$, $A_{12}^{UC}$, $A_{0}^{UL}$ and $A_{0}^{UC}$ in unpolarized 
$e^+e^- \to h_1 h_2 \, X$ processes at BaBar~\cite{Garzia:2012za}. 
The solid lines correspond to the parameters given in Table~\ref{fitpar-poly}, obtained by fitting the $A_{0}$ Belle asymmetry; the shaded area 
corresponds to the uncertainty on these parameters, as explained in the text.
}
\end{center}
\end{figure}

As BABAR data on $A_{12}$ and $A_0$ should be available soon, we 
show in Figs.~\ref{fig:babar-A12-stand} and \ref{fig:babar-A0-poly} 
our expectations, based on our extracted Collins functions. 
Fig.~\ref{fig:babar-A12-stand} shows the expected values of 
$A_{12}^{UL}$, $A_{12}^{UC}$, $A_{0}^{UL}$ and $A_{0}^{UC}$, as
a function of $z_2$ for different bins of $z_1$, using the parameters 
of Table~\ref{fitpar}, obtained by fitting the SIDIS and the $A_{12}$
Belle data with the standard parameterisation.
Fig.~\ref{fig:babar-A0-poly} shows the same quantities using the 
parameters of Table~\ref{fitpar-poly}, obtained by fitting the SIDIS 
and the $A_{0}$ Belle data with the polynomial parameterisation.

    
The Belle (and BABAR) $e^+e^-$ results on the azimuthal correlations of 
hadrons produced in opposite jets, together with the SIDIS data on the 
azimuthal asymmetry $A_{UT}^{\sin(\phi_h + \phi_S)}$, measured by both
the HERMES and COMPASS Collaborations, definitely establish the 
importance of the Collins effect in the fragmentation of a transversely polarised quark. In addition, the SIDIS asymmetry can only be observed   
if coupled to a non negligible quark transversity distribution. 
The first original extraction of the transversity distribution and 
the Collins fragmentation functions~\cite{Anselmino:2007fs, 
Anselmino:2008jk}, has been confirmed here, with new data and a possible 
new functional shape of the Collins functions. The results on the 
transversity distribution have also been confirmed independently in       
Ref.~\cite{Bacchetta:2012ty}.

A further improvement in the QCD analysis of the experimental data, 
towards a more complete understanding of the Collins and transversity 
distributions, and their possible role in other processes, would require 
taking into account the TMD-evolution of $\Delta_T q(x,\kt)$ and 
$\Delta^N\! D_{h/q^\uparrow}(z,\pp)$. Great progress has been recently 
achieved in the study of the TMD-evolution of the unpolarized and Sivers 
transverse momentum dependent distributions \cite{Collins:2011book, 
Aybat:2011zv, Aybat:2011ge, Aybat:2011ta, Anselmino:2012aa}  
and a similar progress is expected soon for the Collins function and the 
transversity TMD distribution~\cite{Bacchetta:2013pqa}.   

\begin{acknowledgments}
Authored by a Jefferson Science Associate, LLC under U.S. DOE Contract 
No. DE-AC05-06OR23177.
We acknowledge support from the European Community under the FP7 
``Capacities - Research Infrastructures" program (HadronPhysics3, 
Grant Agreement 283286). We also acknowledge support by MIUR under 
Cofinanziamento PRIN 2008.
U.D. is grateful to the Department of Theoretical Physics II of the
Universidad Complutense of Madrid for the kind hospitality extended to him 
during the completion of this work.

\end{acknowledgments}

\bibliographystyle{myrevtex} 
\bibliography{sample}

\end{document}